\documentclass[prb,aps,twocolumn,superscriptaddress,showpacs,preprintnumbers,amsmath,amssymb]{revtex4}
\usepackage{color}
\usepackage{graphicx}% Include figure files
\usepackage{epsf, times}
\def\re#1{(\ref{#1})}

\def\textbf#1{\boldsymbol{#1}}

\begin{document}

\title{Electronic and phononic Raman scattering in detwinned
YBa$_2$Cu$_3$O$_{6.95}$ and Y$_{0.85}$Ca$_{0.15}$Ba$_2$Cu$_3$O$_{6.95}$:\\
$s$-wave admixture to the $d_{x^2-y^2}$-wave order parameter}
\author{M.~Bakr}
\affiliation{Max-Planck-Institut~f\"{u}r~Festk\"{o}rperforschung,
Heisenbergstr.~1, D-70569 Stuttgart, Germany}
\author{A.~P.~Schnyder}
\affiliation{Kavli Institute for Theoretical Physics, University of
California, Santa Barbara, CA 93106, USA}
\author{L.~Klam}
\affiliation{Max-Planck-Institut~f\"{u}r~Festk\"{o}rperforschung,
Heisenbergstr.~1, D-70569 Stuttgart, Germany}
\author{D.~Manske}
\affiliation{Max-Planck-Institut~f\"{u}r~Festk\"{o}rperforschung,
Heisenbergstr.~1, D-70569 Stuttgart, Germany}
\author{C.T.~Lin}
\affiliation{Max-Planck-Institut~f\"{u}r~Festk\"{o}rperforschung,
Heisenbergstr.~1, D-70569 Stuttgart, Germany}
\author{B.~Keimer}
\affiliation{Max-Planck-Institut~f\"{u}r~Festk\"{o}rperforschung,
Heisenbergstr.~1, D-70569 Stuttgart, Germany}
\author{M.~Cardona}
\affiliation{Max-Planck-Institut~f\"{u}r~Festk\"{o}rperforschung,
Heisenbergstr.~1, D-70569 Stuttgart, Germany}
\author{C.~Ulrich}
\affiliation{Max-Planck-Institut~f\"{u}r~Festk\"{o}rperforschung,
Heisenbergstr.~1, D-70569 Stuttgart, Germany}

\date{\today}

\begin{abstract}

  Inelastic light (Raman) scattering has been used to study electronic
  excitations and phonon anomalies in detwinned, slightly overdoped
  YBa$_2$Cu$_3$O$_{6.95}$ and moderately overdoped
  Y$_{0.85}$Ca$_{0.15}$Ba$_2$Cu$_3$O$_{6.95}$ single crystals.  In
  both samples modifications of the electronic
  pair-breaking peaks when interchanging the $a$- and $b$-axis were
  observed.  The lineshapes of several phonon modes involving
  plane and apical oxygen vibrations exhibit
  pronounced anisotropies with respect to the incident and scattered
  light field configurations.  Based on a theoretical model that takes
  both electronic and phononic contributions to the Raman spectra into
  account, we attribute the anisotropy of the
  superconductivity-induced changes in the phonon lineshapes to a
  small $s$--wave admixture to the
  $d_{x^2-y^2}$ pair wave--function.  Our theory allows us to
  disentangle the electronic Raman signal from the phononic part and
  to identify corresponding interference terms.  We argue that the
  Raman spectra are consistent with an $s$--wave admixture
  with an upper limit of 20 percent.
\end{abstract}


\maketitle

\section{Introduction}

The symmetry of the pair wavefunction provides important clues to the
mechanism of high temperature superconductivity. Phase sensitive
experiments have shown that the dominant component of the order
parameter exhibits $d_{x^2-y^2}$ symmetry.~\cite{Woll93,Tsue94} In
cuprates of the YBa$_2$Cu$_3$O$_{7}$-type with orthorhombic crystal
symmetry, an $s$--wave admixture resulting in an anisotropy of the
superconducting (SC) energy gap, $2\Delta$, between the $a$-- and
$b$--axes is expected.
%
%on symmetry grounds.
%
A more general trend of increasing $s$-wave admixture with increasing
doping level has also been reported.~\cite{strohm1997, Hiramachi2007,
  nemetschek, masui2003} In YBa$_2$Cu$_3$O$_{7-\delta}$
(YBCO$_{7-\delta}$), such an admixture has been confirmed by
angle-resolved photoemission spectroscopy (ARPES) \cite{Lu01} and flux
measurements on Josephson junctions.~\cite{Kitt06} However, the
interpretation of the ARPES data is complicated because the surface
properties of YBCO$_{7-\delta}$ may differ from those of the
bulk.~\cite{zabolotnyy} The phase-sensitive measurements, on the other
hand, have yielded only lower bounds (9\%) to the $s$-wave admixture.
In addition, phenomenological calculations based on the
two-dimensional Hubbard model \cite{eremin2005,schnyder2006} have
demonstrated that the in-plane anisotropy of the spin fluctuation
spectrum determined by inelastic neutron scattering on detwinned YBCO
\cite{Mook00,Hink04} is consistent with a small isotropic $s$-wave
admixture to the $d_{x^2-y^2}$-pairing symmetry. However, the same
calculations show that the dispersion and spectral weight of the spin
fluctuations are also influenced by anisotropic hopping parameters
that are difficult to determine independently. Moreover, different
interpretations of the neutron data have also been proposed
\cite{Uhrig04,Vojta06}. Further experiments are therefore required to
conclusively establish the magnitude of the $s$-wave contribution to
the SC gap in YBCO$_{7-\delta}$.

Inelastic light (Raman) scattering is a powerful probe of electronic
and lattice vibrational excitations in high-T$_c$
superconductors.~\cite{Hackl,Thom91} Both electronic and phononic
Raman scattering have been applied to study the superconducting gap
anisotropy in YBCO$_{7-\delta}$. The anisotropy can be inferred from
the energies of gap features of the electronic continuum in various
polarization geometries,~\cite{limonov2000,nemetschek} but the
accuracy of this method is limited by overlap with Raman-active
phonons. Possible manifestations of the gap anisotropy and an
anisotropic electron-phonon coupling in the lineshape of a
Raman-active out-of-plane vibration of the in-plane oxygen atoms of
$B_{1g}$ symmetry have also been investigated.~\cite{Limo98}
However, the analysis of these data has been
contested,~\cite{Stro98} and a quantitative estimate of the gap
anisotropy has not been extracted from them.

The present work was in part motivated by a recent theoretical study
that yielded quantitative predictions for the electronic Raman
continua in a $(d_{x^2-y^2}+s)$-wave
superconductor.~\cite{Schnyder2007} In order to enable a detailed
comparison with these predictions, we have performed Raman scattering
measurements of the temperature dependence of the electronic continua
and phonon modes in twin-free, slightly overdoped YBCO$_{6.95}$ and
moderately overdoped Y$_{0.85}$Ca$_{0.15}$Ba$_2$Cu$_3$O$_{6.95}$
(henceforth YBCO$_{6.95}$:Ca) single crystals. We found that
interference between the electronic and phononic scattering channels
imposed severe limitations on our capability to extract information
about the gap anisotropy from the electronic continuum alone, in
accordance with previous work.~\cite{limonov2000,nemetschek} We have
therefore employed a phenomenological model that treats the electronic
and phononic contributions to the Raman signal on equal footing. This
formalism yields predictions for the energy, intrinsic linewidth, and
Fano parameters of phonons coupled to the electronic continuum as a
function of temperature. From a comparison to the experimentally
determined lineshapes of two phonon modes, the $B_{1g}$ vibration of
the in-plane oxygen at 340~cm$^{-1}$ as well as an apical-oxygen
vibration at 501~cm$^{-1}$, we extract an upper bound of 20\% on the
$s$-wave admixture to the SC energy gap. This is consistent with a
10\% to 15\% admixture of $s$-wave to the $d_{x^2-y^2}$ pair-breaking
peak observed in the electronic continuum in the superconducting
state, and it coincides with the lower bound extracted from the
phase-sensitive measurements \cite{Kitt06}. Similar effects have been
recently reported for an organic superconductor with the same point
group as YBa$_2$Cu$_3$O$_{7-\delta}$ ($D_{2h}$).  In this case,
however, the analysis led to the conclusion that only an
isotropic $s$-wave state is present.~\cite{ivanov00}

This paper is organized as follows. In Sec.~\ref{spectra} we
describe experimental details and discuss the raw Raman spectra.  An
introduction to our theoretical model is given in Sec.~\ref{model}.
In particular, we focus on a simultaneous description of both, the
electronic Raman response and the phonon anomalies above and below
$T_c$. Based on this description we are able to analyze the
superconductivity-induced changes in the phonon lineshape in
Sec.~\ref{phonons}.  A summary of our results and a comparison to
prior work is contained in Sec.~\ref{conclusions}.

\section{Experimental details}
\label{spectra}

High quality single crystals of YBCO$_{6.95}$ and YBCO$_{6.95}$:Ca
were grown by the
top-seeded solution growth method.~\cite{Lin_02}
%
%solution method.
%
The orientations of the crystallographic axes were determined by Laue
x--ray diffraction and polarized light microscopy. The samples were
then cut into rectangular shapes of typical size
3$\times$3$\times$1~mm$^3$, annealed at 520$^\circ$C in a flow of
oxygen gas for 150 hours, and quenched in liquid nitrogen in order to
avoid further oxygen diffusion. The magnetization curves of the
YBCO$_{6.95}$ (YBCO$_{6.95}$:Ca) crystal show an onset of $T_c = 92$ K
(75 K). A transition width $\Delta T_c$ of less than 3~K (6~K)
indicates good homogeneity of the samples.  Employing Tallon's
phenomenological expression,~\cite{Tallon}
%$1-T_c/T_{c,max} = 82.6(p-0.16)^2$, where $T_{c,max}$ denotes the
%maximal transition temperature present at optimal doping,
the hole doping level of our samples is estimated as $0.17$ (0.21) per
planar Cu ion, i.e., they are slightly (moderately) overdoped.  The
crystals were detwinned individually using the procedure described in
Refs.  \onlinecite{Voro93} and \onlinecite{Lin_04}.  Raman spectra
measured at different spots of the \textit{ab}-surface of each sample
look basically the same, thus confirming the homogeneity of our single
crystals.

The Raman scattering experiments were performed in backscattering
geometry using a Dilor XY-triple grating Raman spectrometer with a CCD
camera as detector.  For excitation, the 514.5~nm line of an
Ar$^+$/Kr$^+$ mixed gas laser was used. The
resolution of our spectrometers for this experiment was
%
%is
%
about 3~cm$^{-1}$.
%
%Some of the
%spectra presented in this work were also measured using a Labram
%(Jobin Yvon) single-grating Raman spectrometer with a solid-state
%exciting laser ($\lambda$=532 nm).
%
In order to avoid heating of the sample, the power of the incident
laser was kept below 10~mW at the sample surface with a laser spot of
100 $\mu$m in diameter. The direction of the incident laser light was
always
%
%nearly
%
parallel to
the crystallographic $c$--axis. For the selection rule analysis, a
polarizer and an analyzer were placed into the light path before and
after the sample.

YBCO$_{6.95}$ and YBCO$_{6.95}$:Ca are orthorhombic with the space
group $Pmmm$ ($D_{2h}$ symmetry). In polarized Raman scattering
experiments, excitations of $A_g$, $B_{1g}$, $A_g$+$B_{1g}$, and
$A_g$ symmetries are accessible by using $xx$ (or $yy$), $xy$ (or
$yx$), $x'x'$ (or $y'y'$), and $x'y'$ (or $y'x'$) polarizations of
the incident and scattered light fields,
respectively.~\cite{Thom91,Slakey_89} In this notation $x$ and $y$
correspond to the direction of the electric field of light along the
$a$-- and $b$--axes, whereas $x'$ and $y'$ correspond to the
diagonal directions, i.e. $x' \sim x + y$ and $y' \sim x - y$,
respectively.
%
%The beam direction is along the $c$-axis.
%
As
%
%the difference between $a$- and $b$-axis lengths is small, and
%
most previous work on YBCO has been performed
on twinned specimens, we follow previous publications and use the
tetragonal notation of the polarization symmetries throughout this
paper. In this notation, the structure is described in terms of the
closely related tetragonal point group $D_{4h}$, and the excitations
of $A_g$+$B_{1g}$ symmetry are accessible for either $xx$ or $yy$
polarization, $B_{2g}$ for either $xy$ or $yx$ polarization,
$A_g$+$B_{2g}$ for either $x'x'$ or $y'y'$ polarization, and $B_{1g}$
for either $x'y'$ or $y'x'$ polarization of the
incident and scattered electric field vectors.

\begin{figure}
\includegraphics[width=0.98\linewidth]{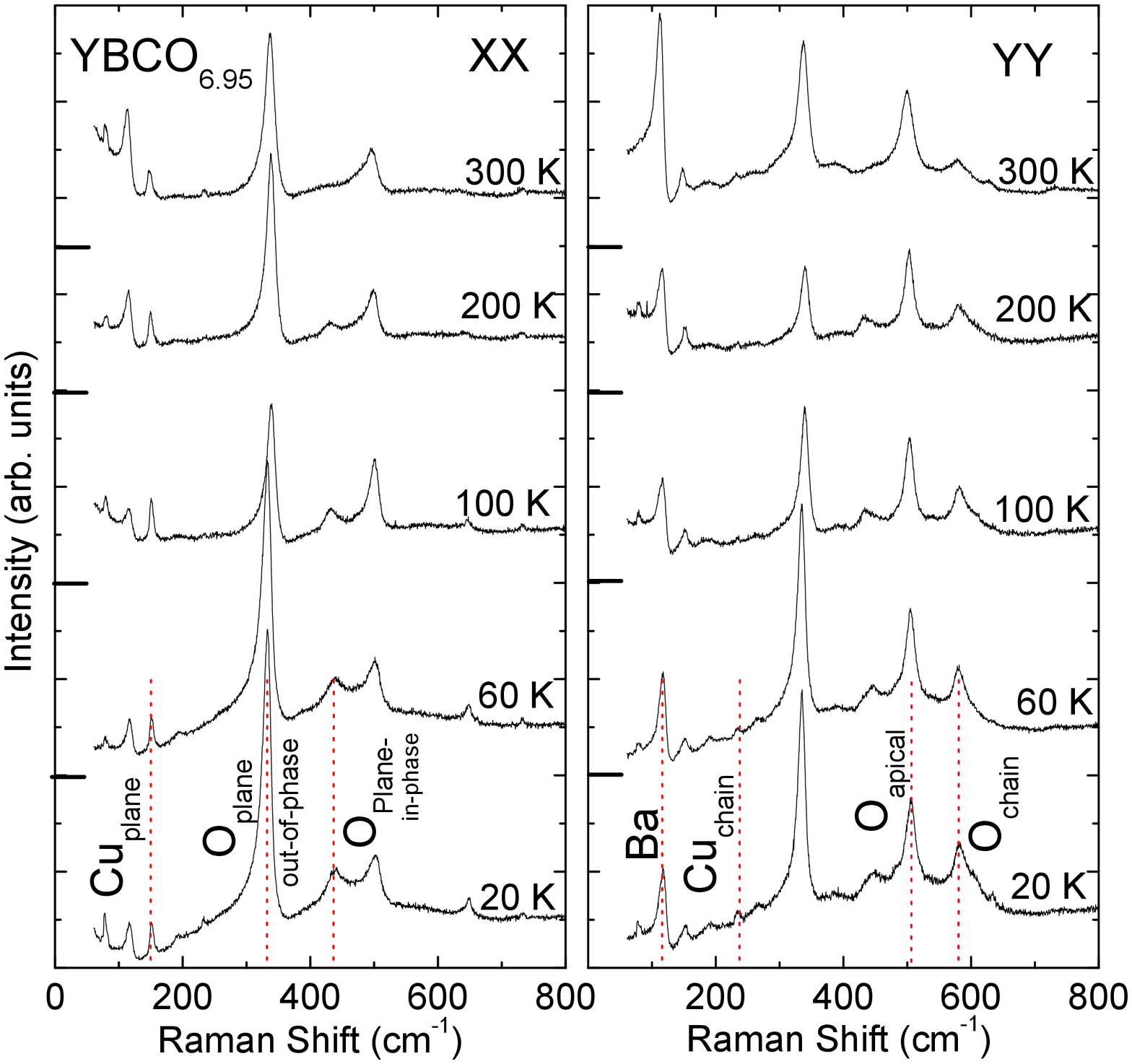}
\caption{\label{opt_spec} Raman spectra of a detwinned slightly
  overdoped YBa$_2$Cu$_3$O$_{6.95}$ single crystal in $xx$ and $yy$
  polarization taken with an Ar$^{+}$ laser line ($\lambda$=514.5 nm).
   The mode assignment corresponds to Refs.
  \protect\onlinecite{Thom88,Thom89,Card99,Wake91,Kaczmarczyk}. The
  spectra were shifted by a constant offset with respect to each
  other. The intensity scales in left and right panels are the same}.
\end{figure}

Figure \ref{opt_spec}
%
%and \ref{ca_spec}
%
show Raman spectra of detwinned YBCO$_{6.95}$
%
%and YBCO$_{6.95}$:Ca
%
crystals measured at various temperatures.  As in other high-$T_c$
superconductors,~\cite{Hewitt_02,Chen_94,Chen_98,Tacon2006,Hackl} the
normal-state spectra exhibit a flat electronic continuum with
superimposed phonons. Below $T_c$, a significant fraction of the
electronic spectral weight is transferred to higher energies,
resulting in a broad pair-breaking peak in the electronic continuum.
In addition, the phonon lineshapes reveal characteristic changes. In
B$_{1g}$ polarization this spectral-weight redistribution is most
pronounced because electronic Raman scattering in this geometry is
sensitive to the maximum of the SC gap along the anti-nodal direction
of the two-dimensional SC gap.  We therefore focus in this work on
this geometry, along with the $xx$ and $yy$ polarization channels that
provide direct information about the in-plane anisotropy of the
electronic response.

%%%%%%%%%%% FIG.1: %%%%%%%%%%%%%%%%%%%
%%%%%%%%%%% FIG.1: %%%%%%%%%%%%%%%%%%%

%%%%%%%%%%%%%%%%%%%%%%%%%%%%%%%%BEGIN FIGURE %%% is now removed %%%
%%\begin{figure}[t]
%%\begin{center}
%%  \includegraphics[width=0.45\textwidth]{fig2.eps}
%%\end{center}
%%\caption{ (color online) Raman spectra of a detwinned
%%  optimally doped Y$_{0.85}$Ca$_{0.15}$Ba$_2$Cu$_3$O$_{6.95}$ sample in $xx$ and
%%  $yy$ polarization ($\lambda$=514.5 nm)}.
%%\label{ca_spec}
%%\end{figure}
%%%%%%%%%%%%%%%%%%%%%%%%%%%%%%%%END FIGURE

Figure \ref{opt_spec} gives an overview of the phonon modes of
YBCO$_{6.95}$ in the $xx$ and $yy$ polarization geometries. As
expected based on a group-theoretical
analysis,~\cite{Thom91,Thom88,Thom89,Card99,Wake91,Kaczmarczyk} five
phonons appear in both polarization channels. The lowest-frequency
phonon at 113 cm$^{-1}$ ($A_{1g}$) originates
predominantly from vibrations of the barium atoms.  The next lowest
frequency phonon at 148 cm$^{-1}$ ($A_{1g}$)
corresponds mainly to the vibrations of copper (Cu2 ions). The
vibrations of the apical oxygen ions (O4) appears at 501 cm$^{-1}$
($A_{1g}$).  The two modes at 340 cm$^{-1}$
($B_{1g}$) and 446 cm$^{-1}$
($A_{1g}$) originate from out-of-phase and in-phase
vibrations of the planar oxygen ions (O2 and O3),
respectively.~\cite{Card99} Two additional defect-induced modes appear
in the $yy$ symmetry at 232 cm$^{-1}$ and at 579 cm$^{-1}$.  They
originate from vibrations of the copper (Cu1) and the oxygen (O1)
ions, respectively, in the Cu--O chains, which are aligned along the
crystallographic $b$--axis.  \cite{Wake91} In YBCO crystals with fully
oxygenated chains, these two modes are Raman forbidden but
infrared--allowed (B$_{1u}$
symmetry).~\cite{Panfilov_98,Bahrs_04,Iliev} They become Raman-active
due to breaking of the translational symmetry by defects, {\it i.e.},
unoccupied oxygen positions in the Cu1-O1
chains.~\cite{Thom88,Thom89,Wake91} The absence of the strong mode at
579 cm$^{-1}$ in the data with $xx$ polarization confirms the high
detwinning ratio of our crystal ($\sim 95$\%).

%
%In Ca-doped YBCO (see Fig.~\ref{ca_spec}), the Ca ions partially
%replace the Y ions at the center of inversion of the YBCO unit cell,
%thus the Raman activity of their vibrations, if any, should be very
%weak.  While the energy of the pair-breaking peak is lower than the
%one in slightly overdoped YBCO$_{6.95}$ (Fig. \ref{opt_spec}), the
%assignment of the phonon modes is identical.  Again, the anisotropy of
%the phonon spectrum confirms the nearly perfect detwinning ratio of
%our YBCO$_{6.95}$:Ca crystal.
%

% Based on the intensity ratio of these phonons, we estimate a
% population ratio of the twin domains of better than 95 \%.  Comment
% to Cardona's corrections: Note that Raman scattering efficiencies
% are certainly $R_{xx}$ \neq R_{yy}$.  Problem: they were not
% calculated by T. Heyen or B. Lederle for this particular mode.
% Therefore, the value of 95\% is an estimated but conservative upper
% limit. The detwinning ration is probably better.

\section{Analysis of Raman spectra}
\label{model}

The measured Raman intensity $I_{\sigma} ( \omega )$ in a given
polarization channel $\sigma$ is related to the imaginary part of
the
%
%electronic
%
response function $ \mathrm{Im} \, \chi_{\sigma} ( \omega )$ via $
I_{\sigma } ( \omega ) = A \left[ 1 + n ( \omega ) \right] \mathrm{Im}
\, \chi_{\sigma } ( \omega ) $, where $n(\omega)$ denotes the Bose
distribution and $A$ is a coupling constant. The Raman response in
cuprate superconductors in the optimally and overdoped regimes
consists of electronic (intraband) and phononic excitations. In order
to disentangle these two contributions it is necessary to employ a
proper fitting procedure.  A phonon interacting with intraband
excitations acquires a renormalized self-energy and exhibits an
asymmetric, Fano-type Raman lineshape.  To describe
such an electron-phonon coupled Raman spectrum the following formula
can be employed
\cite{chen1993,devereaux1995,bock1999,bock1999b,limonov2001a,Letacon07}
\begin{eqnarray} \label{eq: Fano Raman intensity}
&&
\mathrm{Im} \, \chi_{\sigma} ( \omega)
=
 \rho_{\sigma} (\omega) + \frac{ g^2_{\sigma } } {
    \Gamma (\omega) \left[ 1 +
\epsilon^2(\omega) \right] }
\\
&&
\qquad
 \times
\left\{
     S^2(\omega )- 2 \epsilon ( \omega)
    S(\omega) \rho_{\sigma}  ( \omega ) - \rho^2_{\sigma } ( \omega)
  \right\} ,
  \nonumber
\end{eqnarray}
where $\epsilon ( \omega ) = ( \omega^2 - \Omega^2 ) /2 \omega_0
\Gamma ( \omega )$ and $S( \omega ) =S_0 + R_{\sigma}
(\omega )$.  The renormalized phonon frequency and the renormalized
phonon line width are given by $\Omega^2 = \omega_0^2 - 2
\omega_0g^2_{\sigma } R_{\sigma} ( \omega )$, and
$\Gamma ( \omega )= \Gamma_0 + g^2_{ \sigma }
\rho_{\sigma} ( \omega )$, respectively, with the
intrinsic phonon frequency $\omega_0$ and the bare
phonon line width $\Gamma_0$.  The parameter $S_0$ can be expressed in
terms of the electron-phonon coupling $g_{\sigma}$, the Raman phonon
matrix element $T_{\sigma}$, and the Raman electronic matrix element
$\gamma_{\sigma}$, that is, $S_0 = T_{\sigma} /
(\gamma_{\sigma}.g_{\sigma}) $. In our model, $g_{\sigma}$,
$T_{\sigma}$, and $\gamma_{\sigma}$ are assumed to be real and
therefore S$_0$ is real. The Kramers-Kronig related functions
$R_{\sigma} ( \omega)$ and $\rho_{\sigma} (\omega)$ denote real and
imaginary parts of the electronic response function, respectively,
$\chi_{\sigma} (\omega) = R_{\sigma} ( \omega ) + i \rho_{\sigma}
(\omega)$.  While the first term in Eq.~\re{eq: Fano Raman intensity},
$\rho_{\sigma} ( \omega )$, describes the ``bare'' electronic Raman
response, the second term represents the phononic contribution and its
coupling to the electronic background.
We will use expression~\re{eq: Fano Raman intensity} to describe the
coupling of the electronic background to the phonons whose
lineshapes exhibit the clearest manifestations of the
electron-phonon interaction (e.g., the in-plane $B_{1g}$ phonon).
%
% The other phonons, which are located near $115$, $150$, $232$,
% $440$, and $597$ cm$^{-1}$, respectively, can be described by
% Lorentzians with $ I_{ j } ( \omega ) = A_j \Gamma_j/ [ \Gamma_j^2 +
% ( \omega - \Omega_j )^2 ] $.
%
The electronic response function $\chi_{\sigma} ( \omega)$ can either
be computed from a microscopic model~\cite{devereaux1995} or be
determined from a fit to the Raman data using a phenomenological model
function.~\cite{bock1999,bock1999b,limonov2000,limonov2001a}  These
two approaches will be described in the following two subsections.

It is instructive to note that formula \re{eq: Fano Raman intensity}
can be brought into the form of the widely used ``standard'' Fano
profile~\cite{Klein83,limonov2000,Thom91,Thom89}
\begin{eqnarray} \label{eq: fano profile}
I_{\textrm{F} } (\omega)
=
  C_{\textrm{F}  }   ( q + \epsilon)^2 /  ( 1 + \epsilon^2 )  ,
\end{eqnarray}
with the asymmetry parameter
\begin{eqnarray}
q= - S( \omega) / [ \rho ( \omega ) ]
\end{eqnarray}
and $\epsilon = ( \omega - \omega_0 ) /
\Gamma $ as in Eq.~\re{eq: Fano Raman intensity}.
To extract renormalized phonon parameters the Raman spectra are often
fitted using the above
%
%simplified
%
Fano profile, Eq.~\re{eq: fano
  profile}, with the Fano parameter $q$, the
%
%renormalized
%
intrinsic phonon frequency
$\omega_0$, and the renormalized phonon line width
$\Gamma$ (half-width at half-maximum, HWHM) kept frequency
independent.  While such an approach can give valuable insights into
the temperature dependence of the phonon lineshapes, the {\it
  intrinsic} electron-phonon coupling constants and the shape of the
electronic continuum $\chi(\omega)$ in the SC state cannot be
determined.  We will compare the simplified Fano formula \re{eq: fano
  profile} with our generalized theory [Eq.~(\ref{eq: Fano Raman
  intensity})] in Sec.~\ref{phonons}.

%%%%%%%%%%%%%%%%%%%%%%%%%%%%%%%%%%%%%%%%%%
%%%%%%%%%%%%%%%%%%%%%%%%%%%%%%%%%%%%%%%%%%

\subsection{Phenomenological model of the electronic response function}

 %%%%%%%%%%%%%%%%%%%%%%%%%%%%%%%%BEGIN FIGURE 2
\begin{figure}[t]
\begin{center}
\includegraphics[width=0.42\textwidth]{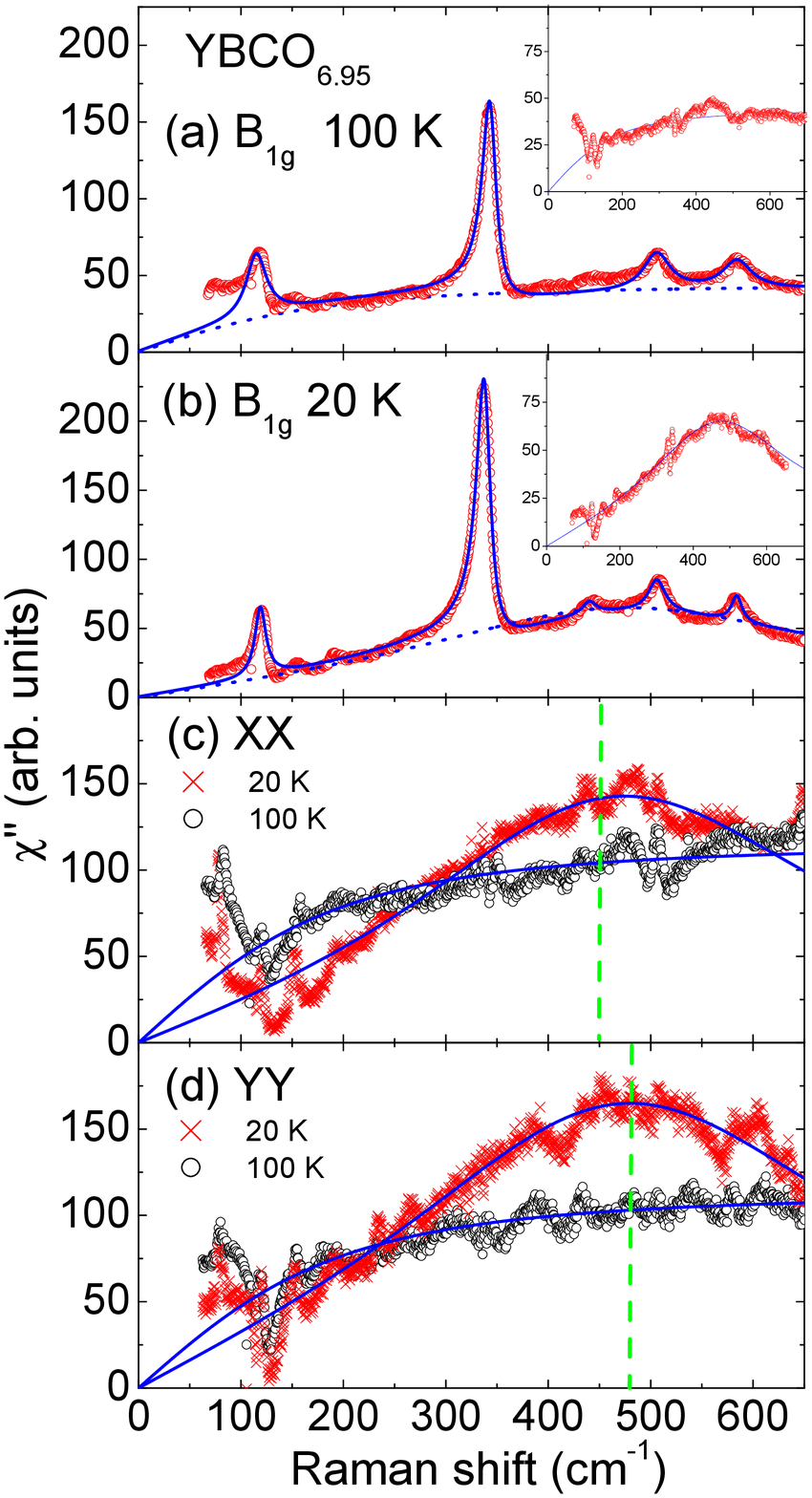}
\end{center}
\caption{ (color online) $B_{1g}$ Raman spectra of YBCO$_{6.95}$
  ($\lambda=514.5$ nm) in the normal state at $T=100$~K (a) and in the
  superconducting state at $T=20$~K (b). Open circles show the
  experimental data, the solid curve the fitting result
  with formula \re{eq: Fano Raman intensity}
  and four Lorentzians for the remaining $A_{1g}$
    phonons.  The insets show the corresponding data after phonon
  subtraction. Panels (c) and (d) show the phonon-subtracted spectra
  for the $xx$ and $yy$ channels, respectively.  The vertical lines
  indicate the maxima of the electronic peak intensity
at 450~cm$^{-1}$ and 480cm$^{-1}$, respectively.}
\label{b1g_spec}
\end{figure}
%%%%%%%%%%%%%%%%%%%%%%%%%%%%%%%%%END FIGURE

To apply
%
%complete
%
the phenomenological model function [Eq.~\re{eq: Fano
Raman intensity}] we need to either assume an expression that
describes the real and imaginary parts of the electronic response
function $\chi(\omega)$, or compute the electronic Raman response
from a microscopic model.  Let us first describe a phenomenological
model function for $\chi (\omega)$. Following along the lines of
Refs.~\onlinecite{bock1999,bock1999b,limonov2001a,limonov2001b} we
express the imaginary part of the Raman efficiency with three terms
\begin{eqnarray} \label{rho model function}
\rho ( \omega )
&=&
C_n \frac{ \omega }{ \sqrt{ \omega^2 + \omega^2_T }}
+
\left[
\frac{C_{1 } }{ 1 + \epsilon^2_{1} ( \omega ) }
- ( \omega \to - \omega )
\right]
\nonumber\\
&& \qquad
-
\left[
\frac{C_{2} }{ 1 + \epsilon^2_{2} ( \omega ) }
- ( \omega \to - \omega )
\right] ,
\end{eqnarray}
where $\epsilon_{1} (\omega ) = ( \omega - \omega_{1} ) / \Gamma_{1} $
and $\epsilon_{2} (\omega ) = ( \omega - \omega_{2} ) / \Gamma_{2 } $.
The first term in Eq.~\re{rho model function} models an incoherent
electronic background, which dominates the response in the normal
state of cuprate superconductors. It is linear in $\omega$ at small
frequencies and becomes constant for large Raman shifts
~\cite{Letacon07}.  The second and third terms are Lorentzians
describing the pair-breaking peak located below $2 \Delta$, and the
suppression of spectral weight at low frequencies, respectively. The
latter is due to the opening of the SC gap. To reduce the number of
free fitting parameters, we set $\omega_2 = \Gamma_2 = ( \omega_1 -
\Gamma_1) / 2$.  The last two terms in Eq.~\re{rho model function}
decrease in intensity and
%
%peak at
%
shift spectral weight to lower frequency as the superconducting
transition temperature $T_c$ is approached from below.  ($\omega
\rightarrow -\omega$) terms, similar to the C$_{1}$ and C$_{2}$ terms
but with $\omega$ replaced by $-\omega$, are essential to achieve the
symmetry requirements for the Raman response. In the normal state, the
electronic response function is entirely described by the incoherent
contribution [first term in Eq.~\re{rho model function}].  The real
part of the electronic response function, $R(\omega)$, is obtained
from the Kramers-Kronig transform of $\rho(\omega)$. Since $R( \omega
)$ renormalizes the phonon frequency $\omega_0$ and the parameter
$S_0$, formula \re{eq: Fano Raman intensity} together with Eq.~\re{rho
  model function} and its Kramers-Kronig transform yield a
self-consistent analysis of the Raman spectra.  To compute the
Kramers-Kronig transform of the incoherent part in Eq.~\re{rho model
  function}, a cut-off frequency $\omega_c$ has to be introduced,
which results in a constant offset in the real part of the electronic
response $R(\omega)$. Provided $\omega_c$ is chosen large enough,
however, this error only leads to negligibly small corrections.

%%%%%%%%%%%%%%%%%%%%%%%%%%%%%%%%%%%%%%%%%%
%%%%%%%%%%%%%%%%%%%%%%%%%%%%%%%%%%%%%%%%%%

%
%We have fitted Eq.~(\ref{eq: Fano Raman intensity})
%with the frequency independent fit parameters $\omega_0$, $\Gamma_0$,
%and $S_0$, as well as the function~\re{rho model function} (and its
%corresponding real part)
%

\begin{figure}[t!]
\begin{center}
\includegraphics[width=0.42\textwidth]{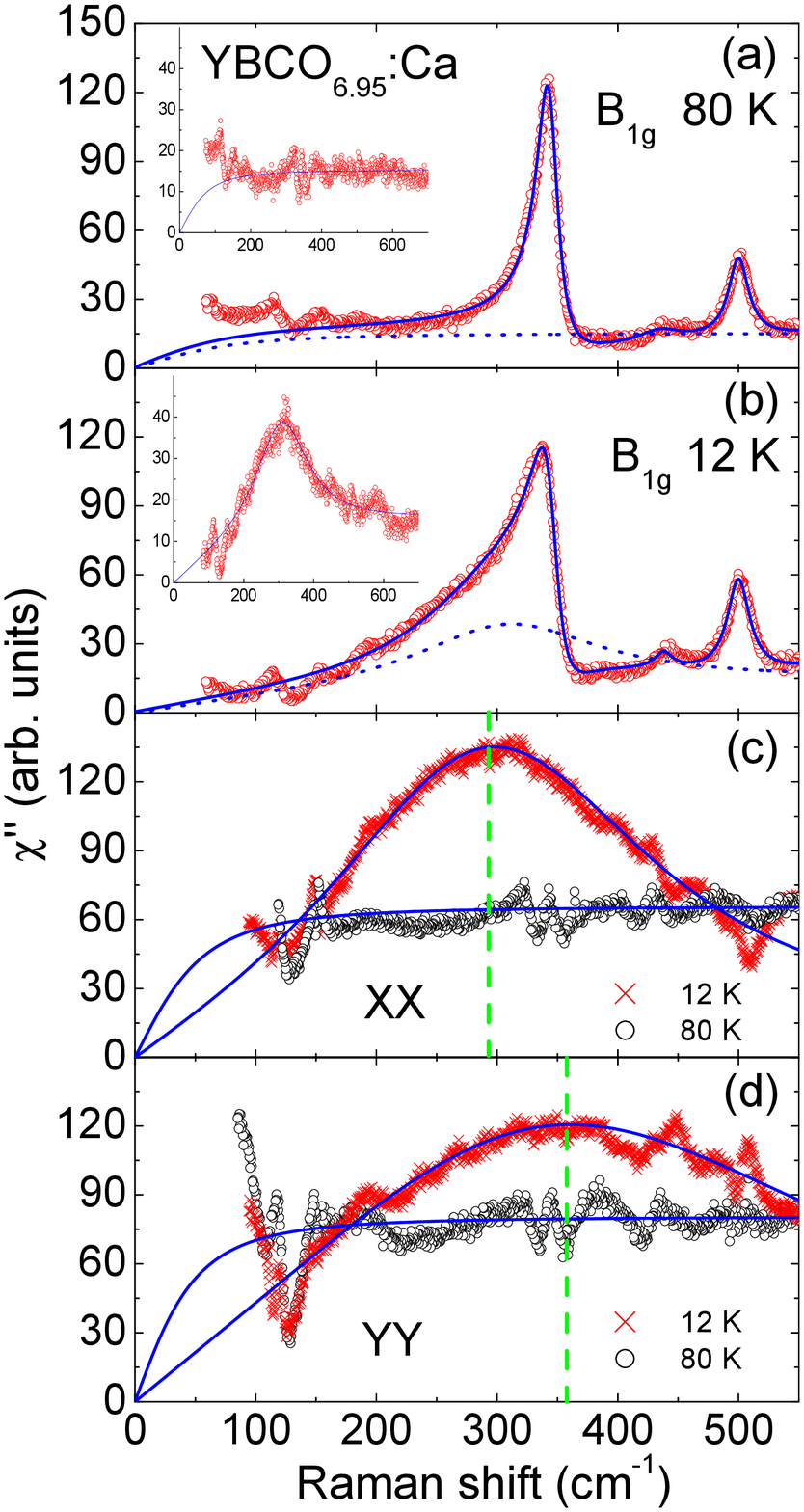}
\end{center}
\caption{(color online) Same as Fig.~\ref{b1g_spec} but for the
  overdoped sample YBCO$_{6.95}$:Ca. The vertical
    lines indicate the maxima of the electronic peak intensity at
    290~cm$^{-1}$ (XX) and 355cm$^{-1}$ (YY), respectively.}
\label{ca_b1gxxyy}
\end{figure}

%
%We have employed a non-linear fit procedure with 10 independent fit
%parameters to the Raman spectra of YBCO$_{6.95}$ and
%YBCO$_{6.95}$:Ca, both in the superconducting and normal state.
%
  We have employed a non-linear fit procedure with ten (six)
  independent fit parameters to the Raman spectra of YBCO$_{6.95}$ and
  YBCO$_{6.95}$:Ca in the superconducting (normal) state.  $C_n$ and
  $\omega_T$ describe the intensity and position of the maximum of the
  normal-state electronic background given by the square root of a
  rational function. In the superconducting state $C_1 / C_2$,
  $\omega_{1,2}$, and $\Gamma_{1,2}$ describe amplitude, position and
  width of a Lorentzian function reflecting the region of the
  pair-breaking peak.  Finally, $\Gamma_0$, $\omega_0$, and $S_0$
  effectively characterize amplitude, width, position, and asymmetry
  of a generalized Fano function (describing the $B_{1g}$ phonon).
  Note that for not too strong frequency dependences of
  $\rho(\omega)$ and $R(\omega)$, the renormalization of the Fano
  formula due to the electronic background can be simply viewed as an
  offset of the parameters entering in the Fano formula [cf. Eqs.\
  (1), (2), and (3)].
%
%%%
%%\begin{table}[t]
%%\begin{ruledtabular}
%%\begin{tabular}{ lrrr }
%% $\hbox{Parameters}$  & $\omega_0$ & $S_0$  & $\Gamma_0^{\rm FWHM}$
%%\\ \hline
%%%
%%YBCO$_{6.95}$ B$_{1g}$ (100K)   & $345.1$  &$3.1$  &$10.8$         \\
%% YBCO$_{6.95}$ XX  (100K)          &$345.3$ &$4.8$  &$11.4$       \\
%% YBCO$_{6.95}$ YY (100K)              &$344.2$ &$5.6$  &$14.4$      \\
%%  YBCO$_{6.95}$ B$_{1g}$ (20K)    &$338.3$  &$3.7$  &$9.6$    \\
%%  YBCO$_{6.95}$ XX (20K)               &$336.7$ &$4.3$   &$12.4$    \\
%% YBCO$_{6.95}$ YY (20K)            &$336.6$ &$5.1$     &$13$    \\
%%
%%%
%%\end{tabular}
%%\end{ruledtabular}
%%\caption{ \label{tab: param} Parameter values of the $B_{1g}$ oxygen
%%vibration extracted from a non-linear least squares fit of the
%%functions described by Eqs. (\ref{eq: Fano Raman intensity}) and
%%(\ref{rho model function}) to the Raman spectra of YBCO$_{6.95}$
%%measured at 20 K and 100 K. The 20 K spectra along $xx$ and $yy$
%%symmetries were measured with the Ar$^{+}$ laser while others were
%%taken with the 532 nm laser line.}
%%%of Figs.~\ref{b1g_spec} and~\ref{ca_b1gxxyy}.
%%\end{table}
%%%
%
%
%
\begin{table}[t]
\begin{ruledtabular}
\begin{tabular}{ l|ccc||c||cc }
YBCO$_{6.95}$ & $S_0$  & $2\Gamma_0$ & $\omega_0$ & $\omega_0$(INS) &
$\omega_0$(sqrt) & $\omega_0$(lin.)
\\ \hline
B$_{1g}$ (100K)  & $1.5$ & $15.3$ & $343.1$ & $343.9$ & $342.8$ & $341.9$ \\
XX  (100K)       & $1.8$ & $12.2$ & $342.9$ & $343.9$ & $339.1$ & $340.7$ \\
YY (100K)        & $1.8$ & $12.2$ & $342.9$ & $343.9$ & $339.5$ & $338.7$ \\
B$_{1g}$ (20K)   & $2.6$ & $14.8$ & $338.6$ & $338.4$ & $335.8$ & $335.7$ \\
XX (20K)         & $4.1$ & $13.2$ & $336.4$ & $338.4$ & $333.0$ & $332.5$ \\
YY (20K)         & $2.1$ & $13.8$ & $336.3$ & $338.4$ & $334.2$ & $334.2$ \\
\end{tabular}
\end{ruledtabular}
\caption{\label{tab: param}Extracted parameter
    values of the $B_{1g}$ oxygen vibration
    in YBCO$_{6.95}$ measured at 20~K and 100~K, respectively. Left
    part: asymmetry parameter $S_0$, intrinsic phonon linewidth (FWHM)
    $2\Gamma_0$, and intrinsic phonon frequency $\omega_0$ using the
    generalized Fano approach of Eq.(\ref{eq: Fano Raman intensity}).
    Right part: $\omega_0$
    extracted with Eq.(\ref{eq: fano profile})
    for different phenomenological electronic
    backgrounds [sqrt = first term in Eq.(\ref{rho model function});
    lin. = linear background
    with offset at $\omega=0$]. For comparison we display in the middle
    part $\omega_0$ obtained from inelastic neutron scattering (INS)
    experiments by Reznik {\it et al.} \protect\cite{reznikprl95}}
\end{table}
The results of this fit procedure are shown in Fig.~\ref{b1g_spec} and
Fig.~\ref{ca_b1gxxyy} together with the experimental data.
Table~\ref{tab: param} lists the corresponding fit parameters.  Note
that on general grounds the extracted $\omega_0$ and $\Gamma_0$
parameter values need to be identical at a given temperature for all
measurements of a given phonon.  In Table~\ref{tab: param} we show,
however, the parameter values that correspond to the best (non-linear
least squares) fit to the data. The differences reflect the error bars
of our procedure.  It is important to emphasize that the Fano profile
of the $B_{1g}$-mode shown in Figs.  \ref{b1g_spec},~\ref{ca_b1gxxyy}
and \ref{sc_b1g_spec}(a) results from the interaction with the
electronic Raman signal; {\it both} electronic and phononic
contributions and their interdependence can be described by Eqs.
(\ref{eq: Fano Raman intensity}) and (\ref{rho model function}). The
final result agrees
%
%very
%
well with the measured data. Thus, our model allows us to some extent
to disentangle the electronic and phononic parts of the Raman
response, and to identify the shape of the electronic background.
However, strictly speaking, only a combination of parameters such as
$g^2\rho$ and $g^2R$ can be extracted.  In the next subsection this
procedure
%
%will be important, in particular,
%
will be improved for the SC state by employing a microscopic description of
the pair-breaking excitations.

 %%%%%%%%%%%%%%%%%%%%%%%%%%%%%%%%BEGIN FIGURE 3
%%%%%%%%%%%%%%%%%%%%%%%%%%%%%%%%%END FIGURE

\subsection{Microscopic description of the superconducting state}

To describe the polarization-dependent electronic response function
$\chi_\sigma (\omega )$ in the superconducting state, we employ a
microscopic model using a realistic tight-binding band structure with
anisotropic hopping parameters and a superconducting gap with a
mixture of $d_{x^2-y^2}$- and $s$-wave symmetries.  Such a microscopic
approach in conjunction with the analysis of the Raman spectra based
on Eq.~\re{eq: Fano Raman intensity} allows us to obtain precise
information about the wavevector dependence of the superconducting
order parameter.  In particular, we are interested in estimating the
magnitude of a possible $s$-wave admixture to the $d_{x^2-y^2}$ pair
wave function.

 The starting point of our calculation is the microscopic model
 introduced in Ref.~\onlinecite{Schnyder2007}. The electronic Raman
 response function in the SC state for a given symmetry
 channel $\sigma$ is described
 by~\cite{Deve94,devereaux1995,Hackl,Schnyder2007}
\begin{eqnarray} \label{eq: chi_micro}
  \chi_{ \sigma } (\omega )
 =
  \left\langle   \gamma_{\sigma }^2
    \theta_{\boldsymbol{k}} ( \omega ) \right\rangle
  -
  \frac{
    \left\langle \gamma_{ \sigma }  \theta_{\boldsymbol{k}}
      ( \omega ) \right\rangle^2
  }
  { \langle \theta_{\boldsymbol{k}} ( \omega )  \rangle } ,
\end{eqnarray}
%
%with
%
$\theta_{\boldsymbol{k}}( \omega )$ being the Tsuneto function.  The
angular brackets denote the average over the Brillouin zone, that is,
\begin{eqnarray}
\label{eq: tsuneto function}
\langle ( \cdots ) \theta_{\boldsymbol{k}} ( \omega ) \rangle
&=&
\frac{1}{V}
\sum_{\boldsymbol{k}} ( \cdots ) \Delta^2_{\boldsymbol{k}} \tanh
\left( \frac{ E_{\boldsymbol{k}} }{ 2 T } \right)
\nonumber\\
& & \hspace*{-10ex}\times
\left(
  \frac{ 1 / E^2_{\boldsymbol{k}} }{ \omega + i \eta + 2 E_{\boldsymbol{k}} }
  -
  \frac{ 1 / E^2_{\boldsymbol{k}} }{ \omega + i \eta - 2 E_{\boldsymbol{k}}  }
\right) .
\end{eqnarray}
The quasiparticle dispersion $ E^{\ }_{\boldsymbol{k}}=
\sqrt{\varepsilon^{2}_{\boldsymbol{k}}+\Delta^{2}_{\boldsymbol{k}}}$
contains an effective one-band description of a single copper-oxygen
plane
\begin{eqnarray}
  \varepsilon^{\ }_{\boldsymbol{k}}&=&
  -
  2t
  \left[
    (1+\delta^{\ }_0)\cos k^{\ }_x
    +
    (1-\delta^{\ }_0)\cos k^{\ }_y
  \right]
  \nonumber\\
  &&
  \qquad -
  4t'\cos k^{\ }_x\cos k^{\ }_y
  -
  \mu ,
\label{eq: dispersion}
\end{eqnarray}
and the superconducting gap
\begin{eqnarray}
\label{eq: s-wavegap}
  \Delta^{\ }_{\boldsymbol{k}}=
  \frac{\Delta_{d}}{2}
  \left(
    \cos k^{\ }_x
    -
    \cos k^{\ }_y
  \right)
  +
  \Delta^{\ }_s .
\end{eqnarray}
For the computation of the electronic Raman response in the
superconducting state, Eq.~\re{eq: chi_micro}, we have assumed the
same band structure parameters as in Ref.~\onlinecite{Schnyder2007}
together with $\Delta_{d} = 30$~meV
%$\Delta_s = 3$~meV,
and $\Gamma_e = 5$~meV.

The Raman vertices $\gamma^{\ }_{\sigma}$ in Eq.~\re{eq: chi_micro}
can be classified according to the irreducible representations of the
symmetry group of the crystal.  Since we shall consider a model with
small distortions, $\delta^{\ }_{0}\ll1$, $\Delta^{\ }_{s}\ll\Delta^{\
}_{d}$, we use the notation corresponding to
tetragonal symmetry, as in Sec.~\ref{spectra} above. We can thus
express, for example, the $B^{\ }_{1g}$ Raman vertex as
\begin{equation}
\gamma^{\ }_{B^{\ }_{1g}\,\boldsymbol{k}}
\propto
t
\left[
\left(1+\delta^{\ }_0\right)\cos k_x
-
\left(1-\delta^{\ }_0\right)\cos k_y
\right] \quad .
\end{equation}

%%%%%%%%%%%%%%%%%%%%%%%%%%%%%%%%%%BEGIN FIGURE 4
\begin{figure}[t]
\begin{center}
\includegraphics[width=0.45\textwidth]{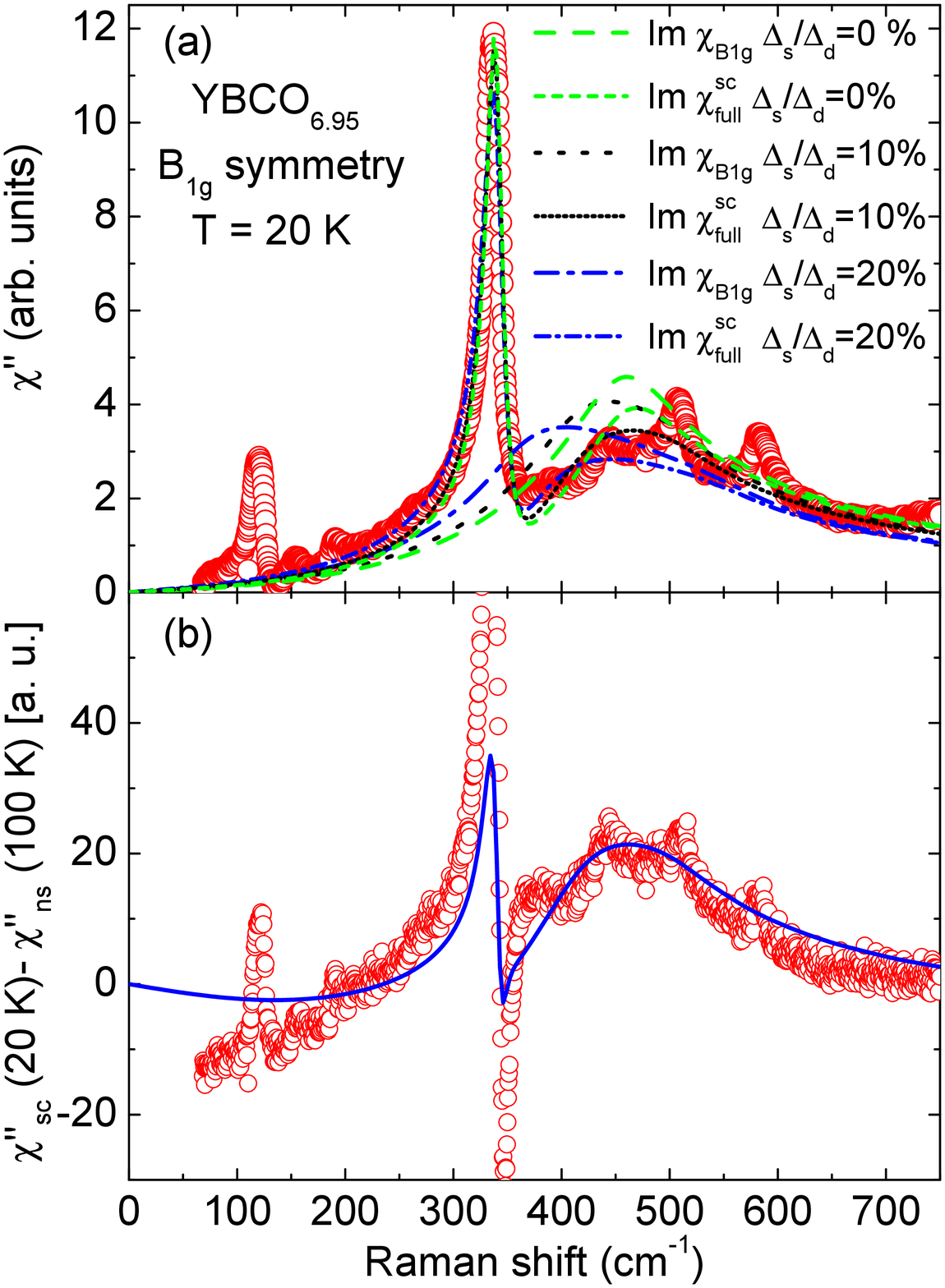}
\end{center}
\caption{ (color online) (a) $B_{1g}$ Raman spectra of YBCO$_{6.95}$
  in the superconducting state at $T=20$~K ($\lambda$=514.5 nm). Open
  circles show the experimental data. The short dashed, short dotted,
  and short dash-dotted curves are the theoretical result obtained
  with Eq.~\re{eq: chi_micro}. The $B_{1g}$ phonon
    was also taken into account. We show results for
    various $s$-wave contributions $\Delta_s / \Delta_d = 0 \,
    (\mbox{dashed}); 0.1 \, (\mbox{dotted}); 0.2 \, (\mbox{dash-dotted})$.
  The dashed, dotted, and dash-dotted curves show the calculated
  imaginary part of the electronic response, $\chi^{\prime
    \prime}_{B_{1g} } ( \omega )$.  (b) Subtracted spectra [20K
  (sc-state)-100K (n-state)] for the B$_{1g}$ polarization channel of
  YBCO$_{6.95}$.  Before subtraction the spectra were divided by the
  Bose factor.  The solid curve depicts the theoretical result
for 10\% $s$-wave contribution.}
\label{sc_b1g_spec}
\end{figure}
%%%%%%%%%%%%%%%%%%%%%%%%%%%%%%%%%%%END FIGURE

To describe the combined electronic and phononic Raman response in
the SC state, we use  Eq.~\re{eq: Fano Raman intensity} together
with the real and imaginary parts of the electronic Raman response
in the SC state, Eq.~\re{eq: chi_micro}. Furthermore, we assume that
the bare fit parameters do not change as we go from the normal state
to the SC state.

%%%%%%%%%%%%%%%%%%%%%%%%%%%%%%%%%%%%%%%%%%%%%
%%%%%%%%%%%%%%%%%%%%%%%%%%%%%%%%%%%%%%%%%%%%%

In Figure \ref{sc_b1g_spec}(a) we show numerical results obtained from
our theoretical model and their comparison with the
data on YBCO$_{6.95}$ obtained in $B_{1g}$-polarization.  The
calculations have been performed at $T=20$ K and for various $s$-wave
contributions.  The best description for both the $B_{1g}$-mode and
the electronic response is found if $\Delta_s$ is assumed to be 10
percent of the maximum of the $d_{x^2-y^2}$-wave gap (short dotted
line); this also accounts for
%
%with
%
the slight shift of the pair
breaking peaks in $xx$ and $yy$ polarizations (Fig.~\ref{b1g_spec} c
and d). Furthermore, our numerical results solely for the electronic
response, i.e. the 2$\Delta$-pair-breaking peak, are also displayed in
Fig. \ref{sc_b1g_spec} (dashed, dotted and dash-dotted lines).  They
are obtained by setting all phononic parts and the corresponding
interference terms in Eq.~(\ref{eq: Fano Raman intensity}) to zero.
Interestingly, we find that the pair-breaking peak shifts to lower
energies with increasing $\Delta_s$, a fact which has been discussed
in a previous paper by Schnyder {\it et al.} \cite{Schnyder2007}. In
addition, the cubic low-energy response, i.e. its
$(\omega/2\Delta_0)^3$-behavior found for
$\Delta_s=0$
changes if
%
%.For the case of
%
$\Delta_s\neq 0$: we obtain linear correction terms which are,
however, proportional to $\Delta_s / \Delta_d$, and thus barely
observable.~\cite{strohm1997}

%%%%%%%%%%%%%%%%%%%%%%%%%%%%%%%%%%BEGIN NEW FIGURE, FIGUR 5
\begin{figure}[t]
\begin{center}
\includegraphics[width=0.45\textwidth]{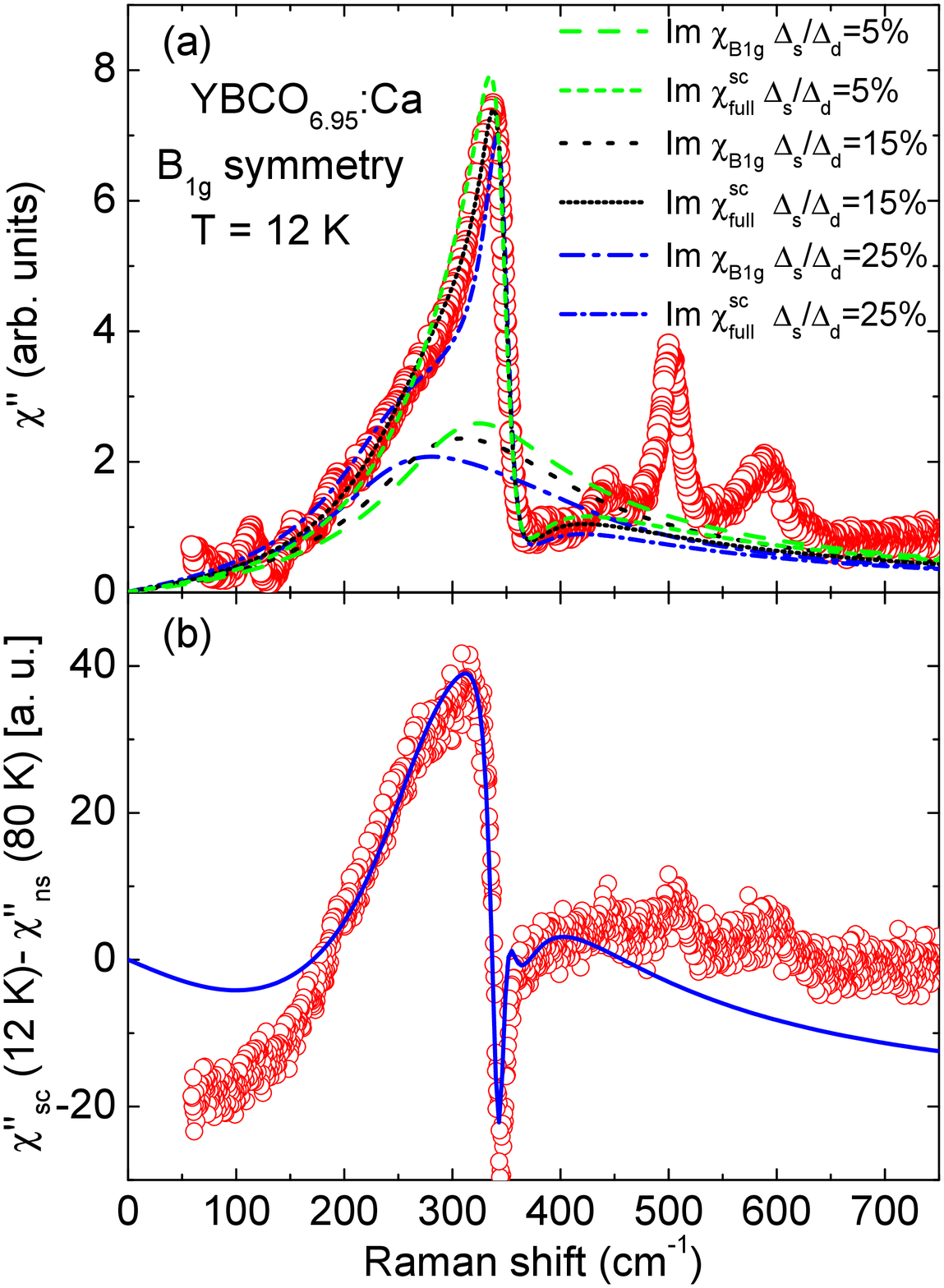}
\end{center}
\caption{(color online) (a) $B_{1g}$ Raman spectra of
    YBCO$_{6.95}$:Ca in the superconducting state at $T=12$~K
    ($\lambda$=514.5 nm). Open circles show the experimental data. The
    short dashed, short dotted, and short dash-dotted curves are the
    theoretical result obtained with Eq.~\re{eq: chi_micro}. The
    $B_{1g}$ phonon was also taken into account. We show results for
    various $s$-wave contributions $\Delta_s / \Delta_d = 0.05 \,
    (\mbox{dashed}); 0.15 \, (\mbox{dotted}); 0.25 \, (\mbox{dash-dotted})$.
    The dashed, dotted, and dash-dotted curves show the calculated
    imaginary part of the electronic response, $\chi^{\prime
      \prime}_{B_{1g} } ( \omega )$.  (b) Subtracted spectra [12K
    (sc-state)-80K (n-state)] for the B$_{1g}$ polarization channel
    of YBCO$_{6.95}$:Ca.  Before subtraction the spectra were divided
    by the Bose factor.  The solid curve depicts the theoretical
    result for 15\% $s$-wave contribution.}
\label{sc_b1g_spec_Ca}
\end{figure}
%%%%%%%%%%%%%%%%%%%%%%%%%%%%%%%%%%%END NEW FIGURE

Figure ~\ref{sc_b1g_spec}(b) shows the differences ($\chi''_S -
\chi''_N$), which were obtained by subtracting the spectra at 20 and
100 K after dividing by the Bose factor.  The solid line is obtained
after subtracting the model function~\re{rho model function} for the
normal state [see Fig.~\ref{b1g_spec}(a)] from the results for the SC
state when $\Delta_s / \Delta_d = 0.1$ [dotted line in
Fig.~\ref{sc_b1g_spec}(a)]. We find that the position of the
pair-breaking peak at $\sim 460$ cm$^{-1}$ (and partly its shape) is
well described by our theory. This confirms that
10\% $s$-wave contribution [short dotted line in
Fig.~\ref{sc_b1g_spec}(a)] yields an excellent description of the
electronic Raman response.  Furthermore, the temperature-dependence of
the related $B_{1g}$ phonon is also reproduced (see
  below). Note that the difference between experimental and
calculated
%
%numerical
%
data at small energies is likely to be due to elastic
impurity scattering which is not taken into account in our theoretical
model.~\cite{remark}

Finally, we return to the moderately overdoped sample.  Raman data on
YBCO$_{6.95}$:Ca above and below $T_c$ in several polarization
geometries are displayed in Fig.
%
%\ref{ca_spec} and
%
\ref{ca_b1gxxyy}.
In the absence of specific information about the electronic band
dispersions of this material, we modeled these data by scaling the
magnitude of the energy gap $\Delta_d$ by the ratio of transition
temperatures, keeping all other model parameters identical to those
used for YBCO$_{6.95}$. The best fit was obtained for $\Delta_s /
\Delta_d = 0.15$ (however the estimated error bars
  are about $\pm 0.05$), slightly larger than the corresponding
quantity in YBCO$_{6.95}$. Note that the quality of the fit is
comparable to the one for YBCO$_{6.95}$, although separate plane and
chain subsystems were not introduced in the analysis.~\cite{Limo05}
The previously observed difference between spectra in $xx$ and $yy$
geometry was confirmed
%
%Fig. \ref{ca_spec},
%
(Fig.~\ref{ca_b1gxxyy}), but our model calculations suggest that this
is a consequence of the $s$-wave admixture to the
gap,~\cite{strohm1997} obviating the need to introduce quantum
interference between scattering from chains and planes.
\cite{Limo05,masui2009}

\section{Temperature dependent phonon lineshapes}
\label{phonons}

\subsection{Anisotropic Fano profile}

%%%%%%%%%%% FIG.6: %%%%%%%%%%%%%%%%%%%
\begin{figure}[t]
\includegraphics[width=0.95\linewidth]{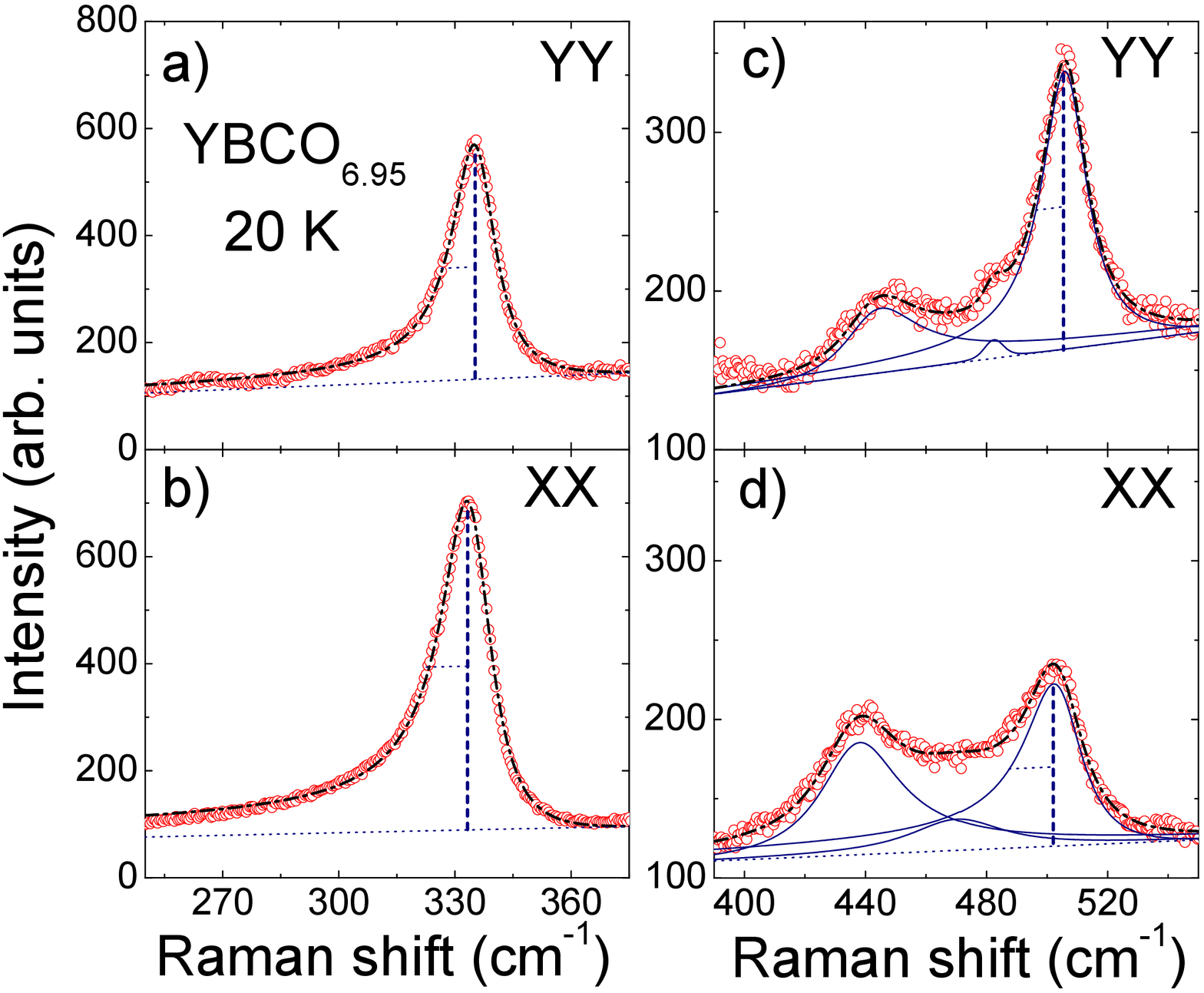}
\caption{\label{fit_phonons} (color online) Fano-analysis
  [Eq.(\ref{eq: fano profile})] of three different
  phonons measured at a temperature of 20 K ($\lambda$=514.5 nm).
  (a,b) The fit of the 340 cm$^{-1}$ mode, which corresponds to the
  out-of-phase vibrations of the planar oxygen, along $yy$ and $xx$
  polarizations, respectively.(c,d) same as (a,b) but for the in-phase
  planar oxygen mode around 440 cm$^{-1}$ and apical oxygen mode at
  501 cm$^{-1}$.  Open circles show the experimental data. The
  dash-dotted lines are the results of the overall fit to the
  experimental data. The solid lines are the results of fits to
  Fano--profiles [Eq.(\ref{eq: fano profile})] as
  described in the text.  The dotted lines correspond to a linear
  background. The intensity units are arbitrary but the same in the
  four vignettes.}
\end{figure}
%%%%%%%%%%%%%%%%%%%%%%%%%%%%%%%%%%%%%%

Several phonons in YBCO display a pronounced
asymmetric lineshape
%
%indicative of
%
suggesting a strong interaction with the electronic
continuum. As shown in Fig. \ref{fit_phonons} the asymmetry is most
pronounced for the 340 cm$^{-1}$ mode.
%
%, but almost all other phonons
%appear to have asymmetric behavior, including the vibration of the
%apical oxygen at 501 cm$^{-1}$.
%
Interestingly,
%
%this
%
the phonon peak reflecting the
  vibration of the apical oxygen at 501cm$^{-1}$ exhibits a strong
asymmetry for a polarization of incident and
  scattered light along the $a$--axis, whereas along the $b$--axis
the phonon appears to be almost symmetric (see Fig.
\ref{fit_phonons}).  Two additional modes are present in this spectral
range.  The mode at 440 cm$^{-1}$ originates from an in-phase
vibration of the oxygen atoms O2 and O3. Additional modes are present
at about 472 cm$^{-1}$ and 480 cm$^{-1}$ for polarizations parallel to
the $a$- and $b$-axis, respectively. As mentioned above, these modes
are Raman forbidden, but correspond to IR allowed
vibrations involving the Cu1-O1 chains.~\cite{Bern02} Due to defects
(oxygen vacancies) in the Cu1-O1 chains they
become Raman active.

Similar to previous temperature dependent Raman experiments on
YBCO$_{7-\delta}$~\cite{Thom88,Alte93} and
HgBa$_2$Ca$_3$Cu$_4$O$_{10+\delta}$,~\cite{Hadj98} we have fitted the
phonons of YBCO$_{6.95}$ by using simple Fano profiles. The solid
lines in Fig. \ref{fit_phonons} are the results of fits to the
experimental data using Eq.~\re{eq: fano profile} , the dash dotted
lines correspond to the resulting fitted lineshape of the entire
spectrum. The calculated profiles agree well with the measured
spectra. In order to obtain estimates of the intrinsic phonon
positions and linewidths, we have corrected the
%
%fitted
%
peak position
and linewidth by the procedure used in Ref. \onlinecite{Stro98}. We
find for the $xx$-symmetry, i.e., polarizations along the $a$-axis:
\begin{eqnarray}
\label{eq: q_xx}
\omega_{xx} =
\omega_{0} + \Gamma / q_{xx} & \ \ {\rm and} \ \ &
\Gamma_{{\rm FWHM}}^{xx} = 2 \Gamma \left| {{(1 + q_{xx}^2)}
    \over {(1 - q_{xx}^2)}} \right| \, .
\end{eqnarray}
$\Gamma_{{\rm FWHM}}^{xx}$ denotes the intrinsic full width at half
maximum (FWHM)\cite{remarkstrohmcardona}. For the
$yy$-polarizations one obtains an analogous equation.

Fits to the simplified Fano profile of Eq.~(\ref{eq: fano profile})
are much less complicated than the fitting procedure to the full
spectrum discussed above. This is a key advantage especially in
situations in which several closely spaced phonons partially overlap
%
%,
%
as is the case, for instance, for the apical-oxygen
vibration in Fig.  \ref{fit_phonons}. The multi-parameter, global fit
[Eq.~(\ref{eq: Fano Raman intensity})]
yields unstable results for these phonons.  It is important to note,
however, that the quantities extracted from simple Fano fits are
renormalized by the electronic response function, and therefore
deviate slightly from the
%
%bare
%
intrinsic phonon frequency $\omega_0$ and the bare
linewidth $\Gamma_0$ of Table~\ref{tab: param}. We will compare the
results of both procedures in detail at the end of the next
subsection.

%{\color{red} Figure ~\ref{xy_correction} shows the
%B$_{1g}$ phonon parameters ($\omega_{p}$ and $\Gamma_{p}$) extracted
%using the standard Fano fit before (Eq.~\re{eq: fano profile}) and
%after asymmetry corrections (Eq.~\re{eq: fano profile} and ~\re{eq:
%q_xx}) in $xx$ and $yy$ polarizations. Using Eq.~\re{eq: fano
%profile} only, one observes an $ab$-differences in the phonon
%parameters outside an error bar of 0.8 cm$^{-1}$. This discrepancy
%is mainly due to the effect of light which produces a difference in
%the asymmetric shape of the phonon peak along $xx$ and $yy$
%polarizations (see fig.~\ref{fit_phonons}). correcting the phonon
%parameters in the $xx$ and $yy$ symmetries as in Eq. ~\re{eq: q_xx}
%brings them to fall together within the error bar. Without such
%corrections, the authors of ref.~\onlinecite{Limo98} claimed that
%the $ab$-anisotropy in the phonon parameters refers to an
%$xy$-discrepancy in the 2$\Delta$-gap leaving their results under
%debate.}
% The resulting phonon energies and linewidths of the
% 340 cm$^{-1}$ and 501 cm$^{-1}$ modes are shown in fig. 6

%
\subsection{Superconductivity-induced changes in the position
and linewidth}

\begin{figure}
\includegraphics[width=0.95\linewidth]{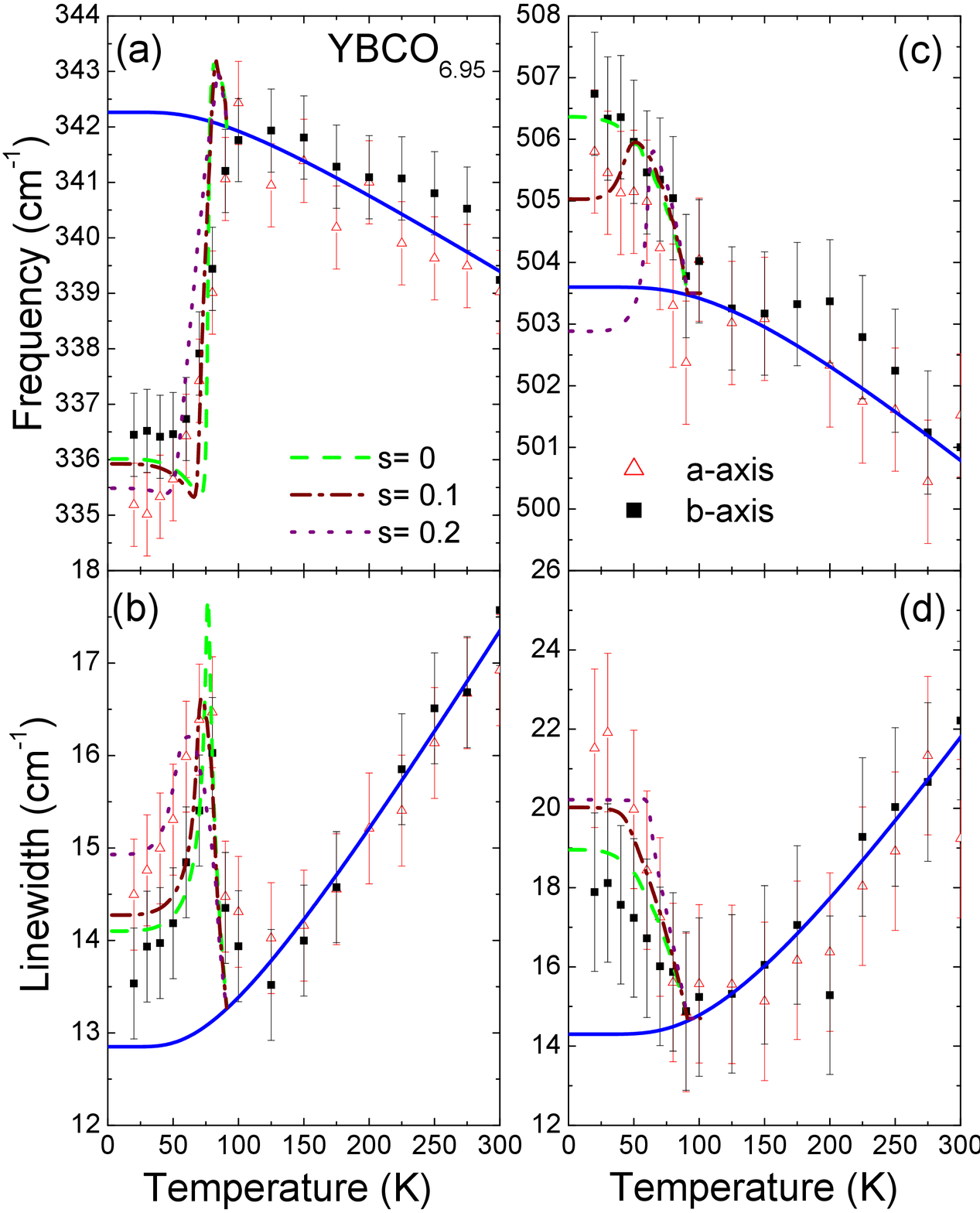}
\caption{\label{sc_shifts} (color online) Frequency and line width
  (FWHM) obtained using Eqs.~(\ref{eq: fano profile})
    and (\ref{eq: q_xx}) versus temperature of the 340 cm$^{-1}$ mode
  and the 501 cm$^{-1}$ mode of YBCO$_{6.95}$ for the $xx$- and
  $yy$-polarization, respectively ($\lambda$=514.5 nm).  Below $T_c$ a
  comparison between calculations for $s = \Delta_s / \Delta_d =0$
  (dashed line), $0.1$ (dash-dotted line), and $0.2$ (dotted line) is
  displayed.  The solid line corresponds to the temperature dependence
  of pure phonon--phonon interaction (interpreted as
  Klemens decay into two phonons of equal frequencies).}
\end{figure}

Figure \ref{sc_shifts} shows the temperature dependence of the
energies and linewidths of two particular phonons, the 340 cm$^{-1}$
and 501 cm$^{-1}$ modes, measured with light polarization along the
crystallographic $a$-- and $b$--axes.  The spectra were taken at
temperatures ranging from 20 to 300 K. The temperature dependence of
the phonon energy and linewidth in the normal state arises from
anharmonic phonon-phonon interactions, i.~e., the decay of a
high-energy optical phonon into two
%
%acoustic
%
 phonons of lower energy
with opposite momenta ~\cite{Klemens66,Mene84}. For simplicity,
assuming the resulting
%
%acoustic
%
 phonons to have the same energy
~\cite{Klemens66,Mene84} and using Bose-Einstein
  statistics, this decay process
%
%is proportional to the acoustic phonon occupation
%number and scales with the Bose-Einstein statistics as
%
leads to $\Gamma_{anh.}(T) = \Gamma_{anh.}^{T=0} ( 1
+ 2 n (\omega_p/2) )$.
%
%This ansatz assumes real transitions. It is highly
%
This process, implying decay
  through real transitions, is strictly valid for the line width, but
has been also used for the frequency shift although,
in this case, virtual transitions also play a role.
In Fig.  \ref{sc_shifts}, both linewidth and peak position were fitted
simultaneously for the temperature range above $T_c$ (solid lines).
Both of these quantities show abrupt changes at the SC transition
temperature due to the opening of the superconducting gap, as
previously observed in YBCO (see for example Refs.
\onlinecite{Thom88,Alte93}) and also other superconducting compounds
like HgBa$_2$Ca$_3$Cu$_4$O$_{10}$ \cite{Hadj98} and
Bi$_2$Sr$_2$CaCu$_2$O$_8$.~\cite{Limonov_02, Mart97}

%%%%%%%%%%% FIG.7: %%%%%%%%%%%%%%%%%%%
%%%%%%%%%%%%%%%%%%%%%%%%%%%%%%%%%%%%%%

Within the error bars of $\sim 0.8$~cm$^{-1}$ in (a)-(c) and
2.0~cm$^{-1}$ in (d), the phonon peak position and linewidth of the
340~cm$^{-1}$ mode is the same for the $xx$ and $yy$ polarizations
(see Fig. \ref{sc_shifts}a and b). Taking the phonon positions and
linewidths in the
%
%normal and
%
SC state
%
%, respectively,
%
as
%
%$\omega_p^{anh}$,
%
$\omega_p^{s}$
%
%, $\Gamma_p^{anh}$,
%
and
$\Gamma_p^{s}$, the maximum change in the phonon position and
linewidth is obtained from $\Delta\omega=
\omega_p^{s}(\tilde{T}) - $
%
%\omega_p^{anh}(T)$
%
$\omega_p(T=100~K)$
and
$\Delta\Gamma=\Gamma_p^{s}(\tilde{T}) - $
%
%\Gamma_p^{anh}(T)$
%
$\Gamma_p(T=100~K)$,~\cite{Hewitt} where $\tilde{T}$
denotes the temperature at which the maximum of the
%
%anomaly
%
SC-induced changes occurs.
The additional softening below $T_c$ due to the electron--phonon
interaction is about -6 cm$^{-1}$. The change in linewidth (FWHM) is
+3 cm$^{-1}$, reflecting a broadening.  For the vibration of the
apical oxygen, i.e. the 501 cm$^{-1}$ mode, the corresponding shift
and broadening are +3 cm$^{-1}$ and +5 cm$^{-1}$, respectively.  The
SC-induced changes for the 340 cm$^{-1}$ and 501 cm$^{-1}$ modes are
in good agreement with previous data obtained on twinned
YBCO.~\cite{Thom88,Alte93,Hewitt}

The SC-induced changes in the phonon linewidth and peak position (for
a phonon labeled as $\mu$) can be related to changes in the phonon
self--energy $\Sigma^{\ }_{\mu}(\omega) = |g^{\mu}|^2 \,\Pi(\omega +
i\eta)$ (with $\eta \to 0$), resulting from the interaction between
the phonons and the electronic system below
$T_c$.~\cite{Deve94,Mart97,Schnyder2007} The induced frequency shifts
$\Delta\omega$ are related to the real part of the phonon polarization
$\Pi$ by~\cite{zeyher,Deve94,Nicol93}
\begin{equation}
\label{eq: phonon self energy_Re}
\frac{\Delta\omega}{\omega} = \frac{1}{N(0)} \, \lambda
\,\mbox{Re}\Pi
,
\end{equation}
while the induced changes in the linewidth can be calculated via
\begin{equation}
\frac{\Delta\Gamma}{\omega} = \frac{1}{N(0)} \,\lambda
\,\mbox{Im}\Pi
.
\end{equation}
Here, $N(0)$ denotes the electronic density of states at the Fermi
level and $\lambda= 2 \sum_{\bf k}\sum_{\mu}\int\frac{d\omega}{\omega}
|g^{\mu}_{{\bf k},0}|^2\, F^{\mu}_{\bf k}(\omega)\delta(\epsilon_{\bf
  k})$ is the dimensionless electron-phonon coupling constant.
$F^{\mu}_{\bf k}(\omega)$ denotes the spectral function for phonon
$\mu$ under consideration.  \cite{mahan_book} Taking into account
screening effects, i.e. the long-range Coulomb force, the self-energy
is given by:~\cite{zeyher,Nicol93,Deve94,Schnyder2007}
\begin{eqnarray} \label{eq: phonon self energy}
\Sigma^{\ }_{\lambda} (\omega)=
-
\left\langle
\left(
g^{\mu}_{\boldsymbol{k},0}
\right)^2
\theta_{\boldsymbol{k}}(\omega)
\right\rangle^{\ }
+
\frac{
\left\langle
g^{\mu}_{\boldsymbol{k},0}
\theta_{\boldsymbol{k}}(\omega)
\right\rangle^{2}
     }
     {
\left\langle
\theta_{\boldsymbol{k}}(\omega)
\right\rangle
     },
\end{eqnarray}
where the angular brackets are defined by Eq.~\re{eq: tsuneto
  function}.  The symmetry of the optical phonons is
reflected in the matrix element
$g^{\mu}_{\boldsymbol{k},0} $.  The electron-phonon
coupling of phonons of $A^{\ }_{1g}$ and $B^{\
}_{1g}$ symmetry are in a first approximation given
by
\begin{equation}
\label{eq: g_constants}
\begin{split}
g^{B^{\ }_{1g} }_{\boldsymbol{k},0}
&=
g^{\ }_{B^{\ }_{1g}} ( \cos k^{\ }_x - \cos k^{\ }_y ) / 2,
\\
g^{A^{\ }_{1g} }_{\boldsymbol{k},0}
&=
g^{\ }_{A^{\ }_{1g}} ( \cos k^{\ }_x + \cos k^{\ }_y ) / 2.
\end{split}
\end{equation}
with the electron-phonon coupling strength $g^{\
}_{B^{\ }_{1g}}$ and $g^{\ }_{A^{\ }_{1g}}$ (Note
  that there is more than one phonon of $A_{1g}$ symmetry).  In
general, the phonons below the energy of the superconducting gap $2
\Delta_{max}$ should soften below T$_{c}$ (i.e.  they should shift to
lower energies), whereas phonons above $2 \Delta_{max}$ should harden.
This is
%
%well
%
confirmed in our experiments (not all data are shown here) and in
previous work (see Refs.  \onlinecite{Limo98}, \onlinecite{Thom88},
and \onlinecite{Alte93}).  The energy of the apical oxygen phonon
(501~cm$^{-1}$ = 62.5 meV) is right at the gap energy and is therefore
sensitive to small changes in the energy of the SC gap.  The 340
cm$^{-1}$ mode is well below the SC energy gap for $T \ll T_c$. With
increasing temperature the energy of the $2 \Delta$ gap shifts to
lower energies and moves through the energy of the 340 cm$^{-1}$ mode.
This explains the maximum of the linewidth at 75 K. A similar behavior
was observed by Limonov $et$ $al.$ \cite{limonov2000}.

Finally, we discuss the role of the $s$-wave contribution to the
superconducting gap which has been introduced in Eq.~(\ref{eq:
  s-wavegap}). Our numerical results obtained with Eqs.~(\ref{eq:
  chi_micro})--(\ref{eq: s-wavegap}) and Eqs.~(\ref{eq: phonon self
  energy_Re})--(\ref{eq: g_constants}) using the fits
reported in Sec.~\ref{model} are also displayed in
Fig.  \ref{sc_shifts}.  We compare $\Delta_s=0$ (dashed line) with
$\Delta_s=3$ meV (dash-dotted line) and $\Delta_s=6$ meV (dotted
line). We find that the results obtained with 20 percent $s$-wave
contribution cannot describe our data. This is clearly visible in
Figs.~\ref{sc_shifts}(a) and (c) which show the SC-induced changes
$\Delta\omega$ in the position of the corresponding phonon. In
particular, for the 501 cm$^{-1}$-mode which is known to be very
sensitive to the superconducting gap,~\cite{Schnyder2007} one predicts
a softening for $\Delta_s=6$ meV, while instead a hardening is
observed.  Therefore, our data imply an upper limit of $\Delta_s /
\Delta_d = 0.2$. The best agreement is obtained for $10\%$ s-wave
admixture.

  At the end of this subsection, we are contrasting our generalized
  Fano theory [see Eq.~(\ref{eq: Fano Raman intensity})] with the
  standard Fano approach described by Eqs.~(\ref{eq: fano profile})
  and (\ref{eq: q_xx}). The main difference between both approaches is
  the theoretical description of the electronic Raman response. While
  in the standard Fano approach the background is assumed either to be
  linear (see Fig.~\ref{fit_phonons}) or to follow a square-root
  behavior \cite{remarktanh} [first term in Eq.~(\ref{rho model
    function})] in both the normal and SC-state, our generalized Fano
  theory is able to take the rearrangement of spectral weight due to
  the opening of the superconducting gap into account. Within our
  microscopic description [Eqs.~(\ref{eq: chi_micro})--(\ref{eq:
    s-wavegap})] it is then possible to determine the ratio $\Delta_s
  / \Delta_d$. Another difference between both Fano theories concerns
  the asymmetry parameter $q$: in our generalized theory it becomes
  $\omega$-dependent which allows a self-consistent description of the
  $B_{1g}$-phonon and the electronic Raman response from 80 to about
  1000~cm$^{-1}$.  Thus, in other words, no specific frequency
  interval close to the phonon peak position [see Fano lineshape
  analysis in Figs.~\ref{fit_phonons}(a) and (b)] needs to be
  selected.

  We have summarized our comparison in Table I in which
  the intrinsic frequency of the $B_{1g}$-phonon extracted from both
  Fano theories is displayed: $\omega_0$ obtained from Eq.~(\ref{eq:
    Fano Raman intensity}) [left part, third column], $\omega_0$
  obtained from Eqs.~(\ref{eq: fano profile}) and (\ref{eq: q_xx})
  either for a square-root [$\omega_0(\mbox{sqrt})$] or linear
  background [$\omega_0(\mbox{lin.})$] (right part). These values have
  to be compared with $\omega_0$ obtained from inelastic neutron
  scattering experiments [$\omega_0(\mbox{INS})$] (middle column). One
  clearly sees that the values for $\omega_0$ obtained from our
  generalized Fano theory are similar to those measured in INS
  experiments. On the other hand, the values extracted from the
  standard Fano approach differ by 1.1 [B$_{1g}$(100K),
  $\omega_0$(sqrt)] to 5.9 [XX(20K), $\omega_0$(lin.)] wavenumbers.
  Note, however, that these values can be substantially modified if a
  specific frequency interval around the peak position of the
  $B_{1g}$-phonon is defined and the corresponding lineshape analysis
  is then restricted to this interval (see Fig.~\ref{fit_phonons}.)
  These improved values are shown in Fig.~\ref{sc_shifts}. In the case
  of the apical oxygen vibration analyzed in
  Figs.~\ref{fit_phonons}(c) and (d), a fit with Eq.~(\ref{eq: Fano
    Raman intensity}) is hardly possible due to the multiple peak
  structure between 400~cm$^{-1}$ and 540~cm$^{-1}$. Therefore, we
  have used Eqs.~(\ref{eq: fano profile}), (\ref{rho model function}),
  and (\ref{eq: q_xx}). On the whole, we find characteristic
  differences between our self-consistent, generalized Fano theory and
  the standard Fano approach, but our conclusions about the ratio
  $\Delta_s / \Delta_d$ are independent of the theory used provided
  that Eqs.~(\ref{eq: fano profile}), (\ref{rho model function}), and
  (\ref{eq: q_xx}) are restricted (or strongly weighted) to an
  interval around the peak position of the phonon being investigated.
\subsection{Temperature dependence of the asymmetry parameter}

%%%%%%%%%%%%%%%%%%%%%%%%%%%%%%%%BEGIN FIGURE 8
\begin{figure}[t]
\begin{center}
\includegraphics[width=0.45\textwidth]{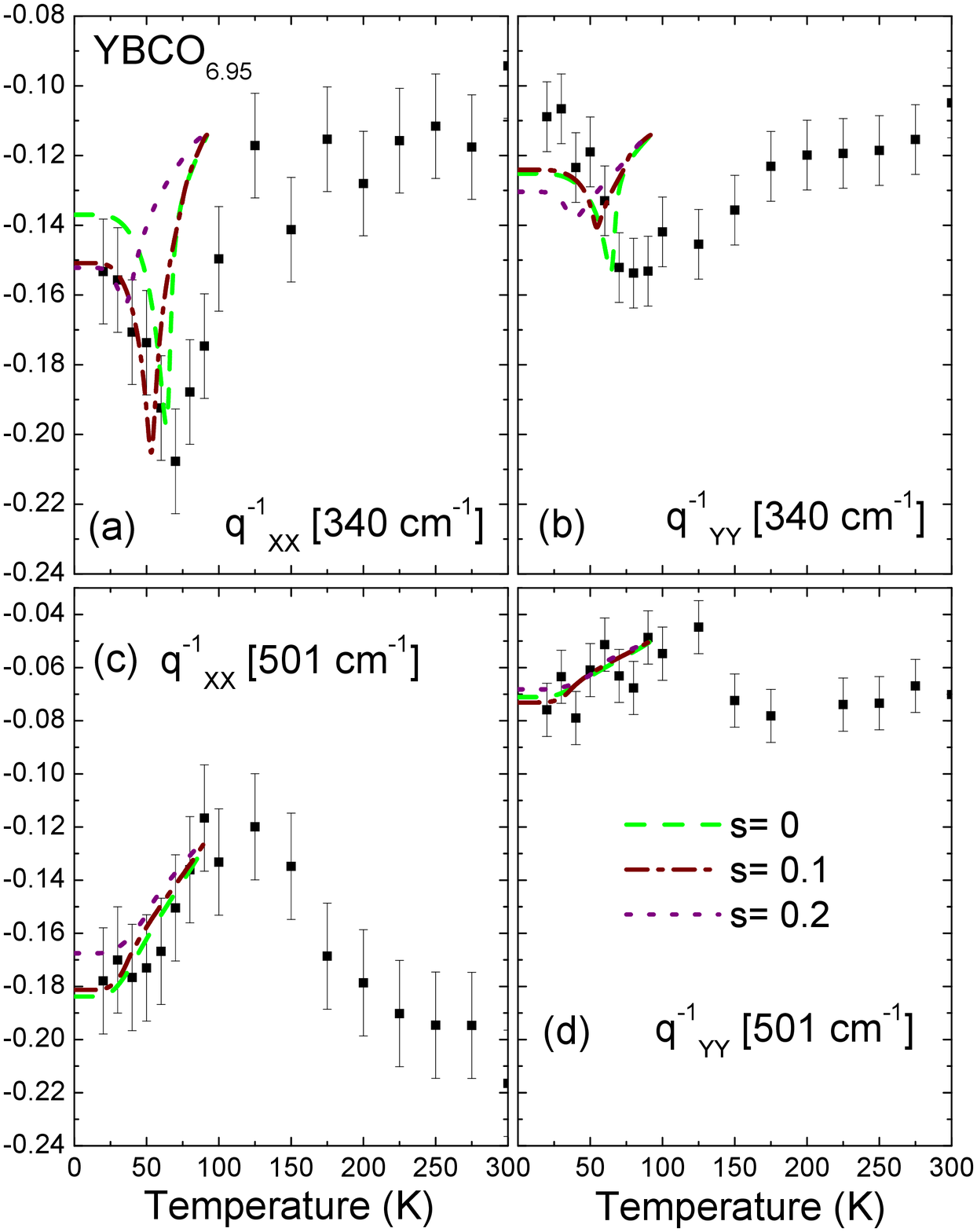}
\end{center}
\caption{ (color online) Fano asymmetry parameter ($\lambda=514.5$
nm) of both polarization channels for the 340 cm$^{-1}$ and 503
cm$^{-1}$ mode of YBCO$_{6.95}$, respectively. Results of the
theoretical calculations for different s-wave admixtures (0, 0.1,
and 0.2) are also shown. } \label{asymmetry}
\end{figure}
%%%%%%%%%%%%%%%%%%%%%%%%%%%%%%%%%END FIGURE

Figure \ref{asymmetry} shows the temperature dependence of the
asymmetry parameters $q^{-1}_{xx}$ and $q^{-1}_{yy}$ in Eq.~(\ref{eq:
  fano profile}) as obtained from the fits of the Fano profile to the
experimental spectra. We discuss these data in the framework of the
approximate expression
\begin{equation}
\label{eq: new_q}
  q \approx \frac{  T_{\sigma} / \gamma_{\sigma}}{g_\sigma \chi''_{\sigma}} ,
\end{equation}
already applied in Raman work on other high $T_{c}$
superconductors.~\cite{Hadj98,Heyen91,Iliev} We first focus
%
%at
%
on temperatures above $T_c$ and note that
$q^{-1}_{xx} \approx q^{-1}_{yy} \sim -0.12$ for the 340 cm$^{-1}$
mode, whereas a pronounced difference in the asymmetry parameters
along in $xx$ and $yy$ geometries is observed for the 501 cm$^{-1}$
mode ($q^{-1}_{xx} \sim - 0.20$, $q^{-1}_{yy} \sim - 0.08$). As
$g_\sigma$ is a materials parameter independent of the light-field
configuration and $\chi''_{\sigma}$ is weakly $\omega$-dependent, the
strong variation of asymmetry parameters must be attributed to
differences in $T_{\sigma} / \gamma_{\sigma}$. Our results are in fair
agreement with the phononic and electronic Raman efficiencies
(proportional to $|T_{\sigma}|^2$ and $|\gamma_{\sigma}|^2$,
respectively) calculated in refs. \onlinecite{Heyen912,Krantz95} using
density functional methods. For instance, the calculated
$\frac{|\gamma/T|_{yy}}{|\gamma/T|_{xx}}$ ($\sim
\frac{q^{-1}_{yy}}{q^{-1}_{xx}}$) in the normal state is about
1.2
%
%1.4
%
and
0.65 for the 340 cm$^{-1}$ and 501 cm$^{-1}$ modes, respectively. The
changes of the asymmetry parameters with temperature in the normal
state may reflect a temperature dependence of
$\gamma_{\sigma}$, {\it i.e.}
%
%and/or
%
parameters in the electronic band
structure as well as that of the anharmonic linewidth, but their
detailed origin cannot be disentangled

Moving to the superconducting state, we note that the only parameter
in Eq.~(\ref{eq: new_q}) expected to change significantly across
$T_c$ is the response function
%
%electronic susceptibility
%
$\chi''_{\sigma}$. The temperature dependence of $q_{xx}^{-1}$ and
$q_{yy}^{-1}$ below $T_c$ therefore also reflects the
\textit{ab}-anisotropy in the 2$\Delta$-gap. Thus we have computed the
temperature evolution of the asymmetry parameters in the
superconducting state, using the same model parameters described in
Sec.~\ref{model}. The reasonable agreement with the data (Fig.
\ref{asymmetry}) demonstrates the self-consistency of our approach.
However, since the superconductivity-induced modification of
$q_{xx}^{-1}$ and $q_{yy}^{-1}$ is relatively subtle (compared to that
observed in the normal state), this analysis does not provide
constraints on the ratio $\Delta_s / \Delta_d$ beyond those already
discussed above.

It is interesting to note that other $A_g$ modes also reveal a
temperature-dependent asymmetry. In particular, we have observed that
the Ba mode near 115~cm$^{-1}$ is symmetric ({\it i.e.} $q^{-1}=0$)
for $T \to 0$ and its asymmetry increases monotonically to
$q^{-1}=0.25$ (not shown).  Its intensity, however, is rather weak
compared with the $B_{1g}$ mode (and this applies even more so to the
Cu mode near 150~cm$^{-1}$), so we have not used these modes further
in our discussion. Their asymmetry has been analyzed in more detail in
Ref.  \onlinecite{ivanov97}, however without considering a possible
$s$-component to the $d_{x^2-y^2}$-wave gap.

\section{Summary and Conclusions}\label{conclusions}

In this work we have revisited the superconductivity-induced effects
on phononic and electronic Raman scattering in detwinned slightly
overdoped YBa$_2$Cu$_3$O$_{6.95}$ and moderately overdoped
Y$_{0.85}$Ca$_{0.15}$Ba$_2$Cu$_3$O$_{6.95}$ single crystals, both from
an experimental and a theoretical point of view.  In particular, we
have performed a detailed study of the in-plane anisotropies in the
electronic continuum and in the phonon lineshapes and assessed their
implications for an $s$-wave admixture to the $d$-wave superconducting
gap. To this end, we developed and applied a formalism that treats
the frequency dependence of both electronic and
phononic Raman scattering on equal footing.  Thus we can disentangle
both parts and clarify the role of interference terms. Due to this
procedure we are able to extract the intrinsic
frequency $\omega_0$ of the phonon position reflecting the intrinsic
electron-phonon interaction.  A comparison shows that
%
%the energy $\hbar\omega_0$
%
$\omega_0$ agrees well with
INS data at the $\Gamma$-point.  The best agreement with the Raman
data was obtained by model calculations based on admixtures of 10\%
and 15\% $\pm$ 5\% $s$-wave contribution for YBa$_2$Cu$_3$O$_{6.95}$
and Y$_{0.85}$Ca$_{0.15}$Ba$_2$Cu$_3$O$_{6.95}$, respectively. This
agrees with values obtained by other experimental methods
\cite{strohm1997,masui2003,Hiramachi2007,nemetschek,Lu01,Kitt06}
and confirms
the previously observed trend \cite{masui2003} of an increase in the
$s$-wave contribution with increasing doping level.  Our data do not
show evidence of the previously reported \cite{Limo05} unusual quantum
interference between electronic Raman scattering from planes and
chains in overdoped Y$_{1-x}$Ca$_{x}$Ba$_2$Cu$_3$O$_{7-\delta}$.

\acknowledgments
We would like to thank R. Zeyher and J.
Unterhinninghofen for helpful discussions and K.
  Syassen for a critical reading of the manuscript. M.B. thanks V.\ Hinkov, R.\
Merkle, and B.\ Baum for their help in sample preparation
and A.  Schulz and H.  Uhlig for technical support.  This work was
partially supported by the International Max Planck Research School
for Advanced Materials (IMPRS).  A.~S. thanks the Swiss NSF for its
financial support and the Max Planck Institute for hospitality.


\begin{thebibliography}{99}
%
\bibitem{Woll93} D.A.~Wollman, D.J.~Van~Harlingen, W.C.~Lee, D.M.~Ginsberg, and A.J.~Leggett,
    Phys. Rev. Lett. {\bf71}, 2134 (1993).
\bibitem{Tsue94} C.C.~Tsuei, J.R.~Kirtley, C.C.~Chi, Lock~See~Yu-Jahnes,
A.~Gupta, T.~Shaw, J.Z.~Sun, and M.B.~Ketchen, Phys. Rev. Lett. {\bf
73}, 593 (1994); C. C.~Tsuei, J. R.~Kirtley , Z. F.~Ren, J. H.~Wang,
H.~Raffy, Z. Z.~
    Li, Nature {\bf 387}, 481 (1997).
\bibitem{strohm1997} T. Strohm and M. Cardona, Solid State Comm. {\bf 104}, 233 (1997); T. Strohm, Ph.D. thesis, University of Stuttgart (1999).
\bibitem{masui2003}T.~Masui, M.~Limonov, H.~Uchiyama, S.~Lee, S.~Tajima, and A.~Yamanaka,
Phys.\ Rev.\ B {\bf 68}, 060506(R) (2003).
\bibitem{Hiramachi2007} T.~Hiramachi, T.~Masui, S.~Tajima, Physica C, {\bf 463}, 89 (2007).
\bibitem{nemetschek} R. Nemetschek, R. Hackl, M. Opel, R. Philipp, M.T. B\'{e}al-Monod,
J.B. Bieri, K. Maki, A. Erb, and E.Walker, Eur. Phys. J. B {\bf 5},
495 (1998).
\bibitem{Lu01} D.H.~Lu, D.L.~Feng, N.P.~Armitage, K.M.~Shen, A.~Damascelli, C.~Kim,
    F.~Ronning, Z.-X.~Shen, D.A.~Bonn, R.~Liang, W.N.~Hardy, A.I.~Rykov, and S.~Tajima,
    Phys. Rev. Lett. {\bf86}, 4370 (2001);  H.~Uchiyama, T.~Masui and
    S.~Tajima, J. Low Temp. Phys. {\bf131}, 287 (2003).
\bibitem{Kitt06} See J.R.~Kirtley, C.C.~Tsuei, A. Ariando, C.J.M.~Verwijs,
S.~Harkema, and H. Hilgenkamp, Nat. Phys. {\bf 2}, 190 (2006), and references therein.
\bibitem{zabolotnyy} V.B.\ Zabolotnyy, S.V.\ Borisenko, A.A.\ Kordyuk, J.\ Geck,
D.S.\ Inosov, A.\ Koitzsch, J.\   Fink, M.\ Knupfer, B.\ B\"uchner, S.-L.\ Drechsler,
H.\ Berger, A.\ Erb, M.\ Lambacher, L.\ Patthey, V.\ Hinkov,
   and B.\ Keimer, Phys. Rev. B {\bf 76}, 064519 (2007).
\bibitem{eremin2005} I.~Eremin and D.~Manske, Phys.~Rev.~Lett. {\bf 94}, 067006 (2005).
\bibitem{schnyder2006} A.~P.~Schnyder, D.~Manske, C.~Mudry, and M.~Sigrist,
    Phys.~Rev.~B {\bf 73}, 224523 (2006).
\bibitem{Mook00}  H.A.~Mook, P. Dai, F.~Dogan, and R.D.~Hunt,
    Nature {\bf 404}, 729 (2000).
\bibitem{Hink04} V.~Hinkov, S.~Pailh\`es, P.~Bourges, Y.~Sidis, A.~Ivanov, A.~Kulakov,
    C.T.~Lin, D.P.~Chen, C.~Bernhard, and B.~Keimer, Nature {\bf 430}, 650 (2004).
\bibitem{Uhrig04} G.~S.~Uhrig, K.~P.~Schmidt, and
M.~Gr\"uninger, Phys. Rev. Lett. {\bf 93}, 267003 (2004).
\bibitem{Vojta06} M.~Vojta, Th.~Vojta, and R.~K.~Kaul, Phys. Rev. Lett. {\bf 97}, 097001
(2006).
\bibitem{Hackl} For a review, see T.P. Devereaux and R. Hackl, Rev. Mod. Phys. {\bf
    79}, 175 (2007).
\bibitem{Thom91} C. Thomsen, in {\it Light Scattering in Solids VI}, edited by
M. Cardona and G. G\"untherodt, Springer Verlag Berlin (1991),
p.285.
\bibitem{limonov2000} % Phonon and electronic Raman scattering in twin-free YBCO
    M.~F.~Limonov, A.~I.~Rykov, S.~Tajima, and A.~Yamanaka,
    Phys.~Rev.~B {\bf 61}, 12412 (2000).
\bibitem{Limo98} M.F.~Limonov, A.I.~Rykov, S.~Tajima, and A.~Yamanaka,
    Phys. Rev. Lett. {\bf 80}, 825 (1998).
\bibitem{Stro98} T.~Strohm, V.I.~Belitsky, V.G.~Hadjiev, and M.~Cardona,
    Phys. Rev. Lett. {\bf 81}, 2180 (1998).
\bibitem{Schnyder2007} A.P.~Schnyder, C.~Mudry, and D.~Manske,
Phys.~Rev.~B {\bf 75}, 174525 (2007).
\bibitem{ivanov00} E. Faulques, V.~G. Ivanov, C. M\'{e}zi\`{e}re, and
P. Batail, Phys. Rev. B {\bf 62}, R9291 (2000).
\bibitem{Lin_02} C.~T. Lin, B. Liang, H.~C. Chen, J. Cryst. Growth
{\bf 237 - 239}, 778 (2002).
\bibitem{Tallon} J.~L. Tallon, C. Bernhard, H. Shaked, R. L. Hitterman,
and J. D. Jorgensen, Phys.~Rev.~B {\bf 51}, 12911 (1995).
\bibitem{Voro93} V.I.~Voronkova and Th.~Wolf, Physica C {\bf 218}, 175 (1993).
\bibitem{Lin_04} C.~T. Lin and A. Kulakov, Physica C {\bf 408-410}, 27
(2004).
\bibitem{Slakey_89} F. Slakey, S.~L. Cooper, M.~V. Klein, J.~P. Rice,
    and D.~M. Ginsberg, Phys. Rev. B {\bf 39}, 2781 (1989).
\bibitem{Hewitt_02} K.~C. Hewitt and J.~C. Irwin,
Phys. Rev. B {\bf 66}, 054516 (2002).
\bibitem{Chen_98} X.~K. Chen, J.~C. Irwin, H.~J. Trodahl, M. Okuya,
T. Kimura, K. Kishio, Physica C {\bf 295}, 80 (1998).
\bibitem{Chen_94} X.~K. Chen, J.~C. Irwin, R. Liang, W.~N. Hardy,
Physica C {\bf 227}, 113 (1994).
\bibitem{Tacon2006} M.~Le Tacon, A.~Sacuto, A.~Georges, G.~Kotliar,
Y.~Gallais, D.~Colson, A.~Forget, Nat. Phys. {\bf 2}, 537 (2006).
\bibitem{Card99} M. Cardona, Physica C {\bf 317-318}, 30 (1999).
\bibitem{Thom88} C. Thomsen, M. Cardona, B. Gegenheimer, R. Liu, and A. Simon,
    Phys. Rev. B {\bf 37}, 9860 (1988);
    B.~Friedl, C.~Thomsen, and M.~Cardona, Phys. Rev. Lett. {\bf 65}, 915 (1990).
\bibitem{Thom89} C.~Thomsen and M.~Cardona in
    {\it Physical properties of high temperature superconductors I},
    ed. by D.M. Ginsberg (World Scientific, Singapore, 1988), p.411.
\bibitem{Wake91} D.R.~Wake, F.~Slakey, M.V.~Klein, J.P.~Rice, and D.M.~Ginsberg,
    Phys. Rev. Lett. {\bf 67}, 3728 (1991).
\bibitem{Kaczmarczyk} C.~Thomsen and G.~Kaczmarczyk in {\it Handbook of vibrational
spectroscopy}, ed. by  J. M.~Chalmers \& P. R.~Griffiths (Wiley, Chichester, 2002), pp. 2651-2669.
\bibitem{Iliev} M. N. Iliev, V.~G. Hadjiev, S. Jandl, D. Le Boeuf,
V.~N. Popov, D. Bonn, R. Liang, W.~N. Hardy, Phys.~Rev.~B {\bf 77},
174302 (2008).
\bibitem{Panfilov_98} A.~G. Panfilov, M.~F. Limonov, A.~I. Rykov,
S. Tajima, and A. Yamanaka, Phys. Rev. B {\bf 57}, R5634 (1998).
\bibitem{Bahrs_04} S. Bahrs, S. Reich, A. Zwick, A.~R. Go\~{n}i, W. Bacsa,
G. Nieva, and C. Thomsen, phys. stat. sol. b {\bf 241}, R63 (2004).
\bibitem{bock1999} % B1g phonon in Pr/Ca doped YBCO
    A.~Bock, S.~Ostertun, R.~Das Sharma, M.~R\"ubhausen, K.-O.~Subke, and C.~T.~Rieck,
    Phys.~Rev.~B {\bf 60}, 3532 (1999).
\bibitem{bock1999b} %R-123
   A.~Bock, Ann.\ Phys.\ (Leipzig) {\bf 8}, 441 (1999).
\bibitem{chen1993} % doping dependence of Raman continua in YBCO
    X.~K.~Chen, E.~Altendorf, J.~C.~Irwin, R.~Liang, and W.~N.~Hardy,
    Phys.~Rev.~B {\bf 48}, 10530 (1993).
\bibitem{devereaux1995} % Theory: B1g Raman phonon in cuprates
    T.~P.~Devereaux, A.~Virosztek, A.~Zawadowski, Phys.~Rev.~B {\bf 51}, 505 (1995).

\bibitem{Letacon07} M.~Le Tacon, A.~Sacuto, Y.~Gallais, D.~Colson, and A.~Forget, % nvestigations
%of the relationship between Tc and the superconducting gap ----
Phys.~Rev.~B {\bf 76}, 144505 (2007).
\bibitem{limonov2001a} %Zn doping effect on Phonons in YBCO
    M.~Limonov, D.~Shantsev, S.~Tajima, and
    A.~Yamanaka, Phys.~Rev.~B {\bf 65}, 024515 (2001).
\bibitem{Klein83} M. V. Klein, in {\it Light Scattering in Solids I}, edited by
    M. Cardona, Springer Verlag Berlin (1983), p. 169.
\bibitem{limonov2001b} % Zn doping in YBCO --- separation of elect and phononic parts
    M.~Limonov, D.~Shantsev, S.~Tajima, and A.~Yamanaka,
    Physica C {\bf 357-360}, 265 (2001).
\bibitem{Deve94} T.P.~Devereaux, Phys. Rev. B {\bf 50}, 10287 (1994).
  %  T.P.~Devereaux, A.~Virosztek, and A.~Zawadowski, Phys. Rev. B {\bf 51}, 505 (1995).
%\bibitem{Card00} M. Cardona in {\it Raman Scattering in Material Science},
%    ed. by W. H. Weber and R. Merlin (Springer Verlag, Berlin 2000).
\bibitem{remark} We emphasize that the low-energy power laws in
  Figs.~\ref{b1g_spec} and Fig.~\ref{sc_b1g_spec}(a) are changed in the
  presence of an $s$-wave contribution while there exists no power law
  for the substracted data in Fig.~\ref{sc_b1g_spec}(b).
\bibitem{Limo05} T.~Masui, M.~Limonov, H.~Uchiyama, S.~Tajima, and A.~Yamanaka,
    Phys. Rev. Lett. {\bf 95}, 207001 (2005); M. Limonov, T. Masui, H. Uchiyama,
S. Lee, S. Tajima, and A. Yamanaka, Physica C {\bf 392-396}, 53
(2003).
\bibitem{masui2009} T.~Masui, T.~Hiramachi, K.~Nagasao, and S.~Tajima,
Phys. Rev. {\bf 79}, 014511 (2009).
\bibitem{Bern02} C. Bernhard, T. Holden, J. Humlicek, D. Munzar, A. Golnik,
    M. Kl\"aser, Th. Wolf, L. Carr, C. Homes, B. Keimer, and M. Cardona,
    Solid State Commun. {\bf 121}, 93 (2002).
\bibitem{Alte93} E.~Altendorf, X.K.~Chen, J.C.~Irwin, R.~Liang, and W.N.~Hardy,
    Phys. Rev. B {\bf 47}, 8140 (1993); E.~Altendorf, J.~Chrzanowski,
    J. C.~Irwin, A.~O'Reilly, and W. N.~Hardy, Physica C {\bf 175}, 47 (1991).
\bibitem{Hadj98} V.G.~Hadjiev, Xingjiang Zhou, T.~Strohm, M.~Cardona,
    Q.M.~Lin, and C.W.~Chu, Phys. Rev. B {\bf 58}, 1043 (1998).
\bibitem{remarkstrohmcardona} Note that the derivation of Eq.~(\ref{eq: q_xx})
assumes that the Raman intensity $I(\omega)$ can be described by
Eq.~(\ref{eq: fano profile}) and an $\omega$-independent background.
On the other hand, if $I(\omega)=\frac{(q+\epsilon)^2}{1+\epsilon^2} +
\alpha\,\omega + \beta$ (with $\alpha \neq 0$) is assumed, one obtains
higher correction terms for $q > 1$: $\omega_{xx} - \omega_0 =
\Gamma / q_{xx} + \frac{1}{2}\frac{1+q_{xx}^2}{q_{xx}^4}\,\Gamma^2\,\alpha
+ ...$. Since, however, the slope $\alpha$ is relatively small, those
higher corrections are of the order of 2\% for the $B_{1g}$-phonon,
for example. Therefore, it is reasonable to work with Eq.~(\ref{eq: q_xx}).
%
\bibitem{Klemens66} P.~G.~Klemens, Phys. Rev. {\bf 148},
845 (1966).
\bibitem{Mene84} J.~Men\'{e}ndez and
M.~Cardona, Phys. Rev. B {\bf 29}, 2051 (1984).
\bibitem{Limonov_02} M. Limonov, S. Lee, S. Tajima, and
    A. Yamanaka, Phys. Rev. B {\bf 66}, 054509 (2002).
\bibitem{Mart97} A.A.~Martin, J.A.~Sanjurjo, K.C.~Hewitt, X.Z.~Wang,
    J.C.~Irwin, and M.J.G.~Lee, Phys. Rev. B {\bf 56}, 8426 (1997).
\bibitem{Hewitt}K.~C. Hewitt, X.~K. Chen, C. Roch, J. Chrzanowski, J. C. Irwin,
    E.~H. Altendorf, R. Liang, D. Bonn, and W.~N. Hardy,
    Phys. Rev. B {\bf 69}, 064514 (2004).
\bibitem{zeyher} R. Zeyher and G. Zwicknagl, Z. Phys. B {\bf 78}, 175 (1990).
\bibitem{Nicol93} E. J. ~ Nicol, C. ~Jiang, and J. P.~ Carbotte,  Phys. Rev. B {\bf 47}, 8131 (1993).
\bibitem{mahan_book} G.~D. Mahan, {\it Many-Particle Physics}, Plenum
  Press, New York (1981).
\bibitem{reznikprl95} D. Reznik, B. Keimer, F. Dogan, and I.~A. Aksay,
Phys. Rev. Lett. {\bf 75}, 2396 (1995).
\bibitem{Heyen91} E.~T. Heyen, M. Cardona, J. Karpinski, E. Kaldis, and S. Rusiecki,
    Phys. Rev. B {\bf 43}, 12958 (1991).
\bibitem{Heyen912} E.T.~Heyen, Ph.D. Thesis, University of Stuttgart, Germany
(1991).
\bibitem{Krantz95} M.~C.~Krantz, I.~I.~Mazin,
D.~H.~Leach, W.~Y.~Lee, and M.~Cardona, Phys. Rev. B {\bf 51}, 5949
(1995).
\bibitem{remarktanh} This is similar to a $\tanh(\omega/4T)$-behavior
of the electronic background assumed in Ref.\protect\onlinecite{bock1999b}.
\bibitem{ivanov97} E. Faulques and V.~G. Ivanov, Phys. Rev. B {\bf 55}, 3974
(1997).





%%%%%%%%%%%%%%%%%%%%%%%%%%%%%%%%%%
%
%\bibitem{Coop88} S.L.~Cooper, F.~Slakey, M.V.~Klein, J.P.~Rice, E.D.~Bukowski, and D.M.~Ginsberg,
%    Phys. Rev. B {\bf 38}, 11934 (1988).
%\bibitem{Tran95} J.M.~Tranquada, B.J.~Sternlieb, J.D.~Axe, Y.~Nakamura, and S.~Uchida,
%    Nature {\bf 375}, 561 (1995).
%\bibitem{Cheo91} S-W.~Cheong, G.~Aeppli, T.E.~Mason, H.A.~Mook, S.M.~Hayden, P.C.~Canfield, Z.~Fisk,
%    K.N.~Clausen and J.L.~Martinez, Phys. Rev. Lett. {\bf 67}, 1791 (1991);
%    T.~Mason, G.~Aeppli, and H.A.~Mook, Phys. Rev. Lett. {\bf 68}, 1414 (1992).

%\bibitem{Lube04} S. V. Lubenets, V. D. Natsik, and L. S. Fomenko,
%    Low Temp. Phys. {\bf 30}, 345 (2004).
%\bibitem{Gall02} Y. Gallais, A. Sacuto, P. Bourges, Y. Sidis,
%A. Forget, and D. Colson, Phys. Rev. Lett. {\bf 88}, 177401 (2002).
%\bibitem{Limo00-1} M.F. Limonov, S. Tajima, and A. Yamanaka,
%    Phys. Rev. B {\bf 62}, 11859 (2000).
%\bibitem{Kran95} M. Krantz and M. Cardona, J. Low Temp. Phys. {\bf 99}, 205 (1995).
%
%
%% Raman scattering, detwinned YBCO
%
%\bibitem{Limo98-2} M.F. Limonov, A.G. Panfilov, A.I. Rykov, S. Tajima,
%and A. Yamanaka,    J. Phys. Chem. Solids {\bf 59}, 1997 (1998).
%\bibitem{Limo98-r} M.F. Limonov, A.I.\ Rykov, S.\ Tajima, and A.\ Yamanaka,
%    Phys. Rev. Lett. {\bf 81}, 2181 (1998).
%\bibitem{Fain00} A. Fainstein, B. Maiorov, J. Guimpel, G. Nieva, and E. Osquiguil,
%    Phys. Rev. B {\bf 61}, 4298 (2000).
%
%%%%%%%%%%%%%%%%%%%%%%%%%%%%%%%%%%%%%%%

%
\end{thebibliography}
\end{document}